\documentclass[journal,twoside,web]{ieeecolor}
\usepackage{generic}
\usepackage{amssymb}
\usepackage{hyperref}
\usepackage{amsmath,bm}
\usepackage{algorithm}
\usepackage{algpseudocode}
\usepackage{cite}
\usepackage{subfig}
\usepackage{tabularx}
\usepackage{tabulary}
\usepackage{mathrsfs}
\usepackage{graphicx}
\usepackage{epstopdf}
\usepackage{multirow}
\usepackage{makecell}
\usepackage{booktabs}
\usepackage{url}
\usepackage[dvipsnames]{xcolor}
\hypersetup{hidelinks}
\definecolor{IEEEBlue}{RGB}{0,138,218}

\begin{document}
\title{Sparse Bayesian Correntropy Learning \\ for Robust Muscle Activity Reconstruction \\ from Noisy Brain Recordings}
\author{Yuanhao~Li,
Badong~Chen,~\IEEEmembership{Senior~Member,~IEEE,}
\\
Natsue~Yoshimura,
Yasuharu~Koike,
and~Okito~Yamashita
\thanks{This work was supported in part by Japan Society for the Promotion of Science (JSPS) KAKENHI under Grants 19H05728, 20H00600, and 23H03433, in part by Innovative Science and Technology Initiative for Security under Grant JPJ004596 ATLA, in part by Moonshot Program 9 under Grant JPMJMS2291, and in part by the National Natural Science Foundation of China under Grants U21A20485 and 62311540022. \emph{(Cor- responding author: Yuanhao Li.)}}
\thanks{This work involved human subjects or animals in its research.Approval of all ethical and experimental procedures and protocols was granted by the Ethics Committee of National Center of Neurology and Psychiatry.}
\thanks{Yuanhao Li and Okito Yamashita are with the Center for Advanced Int- elligence Project, RIKEN, Tokyo 103-0027, Japan, and also with the Br- ain Information Communication Research Laboratory Group, Advanced Telecommunications Research Institute International, Kyoto 619-0237, Japan. (correspondence e-mail: yuanhao.li@riken.jp)}
\thanks{Badong Chen is with the Institute of Artificial Intelligence and Roboti- cs, Xi'an Jiaotong University, Xi'an 710049, China.}
\thanks{Natsue Yoshimura is with the School of Computing, Tokyo Institute of Technology, Yokohama 226-8503, Japan.}
\thanks{Yasuharu Koike is with the Institute of Innovative Research, Tokyo Ins- titute of Technology, Yokohama 226-8503, Japan.}
\thanks{The code is available at \href{https://sites.google.com/view/liyuanhao/code}{\textcolor{IEEEBlue}{\textit{https://sites.google.com/view/liyuanhao/code}}}.}}
	
\maketitle
\begin{abstract}
Sparse Bayesian learning has promoted many effective frameworks for brain activity decoding, especially for the reconstruction of muscle activity. However, existing sparse Bayesian learning mainly employs Gaussian distrib- ution as error assumption in the reconstruction task, which is not necessarily the truth in the real-world application. On the other hand, brain recording is known to be highly noisy and contains many non-Gaussian noises, which could lead to significant performance degradation for sparse Bayesian learning method. The goal of this paper is to propose a new robust implementation for sparse Bayesian learning, so that robustness and sparseness can be realized simultaneously. Motivated by the great robustness of maximum correntropy criterion (MCC), we proposed an integration of MCC into the sparse Bayesian learning regime. To be specific, we derived the explicit error assumption inherent in the MCC and then leveraged it for the likelihood function. Meanwhile, we used the automatic relevance determination (ARD) technique for the sparse prior distribution. To fully evaluate the proposed method, a synthetic dataset and a real-world muscle activity reconstruction task with two different brain modalities were employed. Experimental results showed that our proposed sparse Bayesian correntropy learning framework improves significantly the robustness in a noisy regression task. The proposed method can realize higher correlation coefficient and lower root mean squared error in the real-world muscle activity reconstruction tasks. Sparse Bayesian correntropy learning provides a powerful tool for neural decoding which can promote the development of brain-computer interfaces.
\end{abstract}

	
\section{Introduction}
\label{sec:introduction}
\IEEEPARstart{B}{rain}-computer interface (BCI) has been revealing hopeful and promising prospects for paralyzed and amputated people towards their rehabilitation and communication \cite{chaudhary2016brain,kawala2021summary}. In particular, previous studies have demonstrated the capability of BCI to reproduce the kinetic functions \cite{cervera2018brain,wang2023comprehensive}, which would help neurological recovery for post-stroke motor rehabilitation and develop the assistive prosthesis for motor-impaired people. The crucial procedure for a motion-related BCI is to accurately decode the movements from simultaneous brain activity, which is usually realized by two different strategies. The first scheme is motor imagery \cite{arpaia2022successfully}, a classification scenario with pre-defined labels. However, the discrete label of motor imagery may result in an inadequate capability for the reproduction of complicated motion with only the predetermined motion intention. Another strategy is to reconstruct the continuous movements, which can reflect the motion in real time and thus exhibits larger potential to advance the development of motion-related BCI system \cite{wang2023comprehensive}.

Electromyography (EMG) signal directly reflects the muscle activation, which comprises the information for related motion \cite{bi2019review}, and can be employed to restore the torques and trajectories of muscle activity \cite{koike2024motion}, and to control a prosthetic hand \cite{qin2021multi,qin2022cw}. Previous studies have investigated the decoding of EMG-based muscle activity by different brain recording methods \cite{tam2019human,umeda2019decoding,wang2023eeg,das2023hierarchical,ganesh2008sparse}, including the widely utilized electroencephalogram (EEG) and functional magnetic resonance imaging (fMRI). Further, some have employed the multimodal source imaging methods which could merge the merits of EEG and fMRI, thus realizing better motion decoding performance by the estimated cortical current sources \cite{toda2011reconstruction,yoshimura2012reconstruction,yoshimura2017decoding,mejia2018decoding,sosnik2021reconstruction}.

The regression algorithm plays a vital role for reconstructing muscle activities, which predicts the continuous response from brain recordings. Although recent advances have utilized deep learning methods for brain activity decoding \cite{altaheri2023deep}, conventional linear regression models are still popularly utilized for motion decoding from neural signals in recent studies, e.g., \cite{umeda2019decoding,sosnik2021reconstruction,das2023hierarchical}, which is also mentioned in a recent review paper \cite{wang2023comprehensive}. This is mainly because the deep learning methods need copious data for training while collecting a large amount of brain recordings is not always feasible. Moreover, linear model can improve the interpretability for the neurological patterns \cite{wang2023eeg}. In particular, to alleviate the overfitting problem with a small dataset, sparse Bayesian learning has been widely leveraged for brain activity analysis \cite{ganesh2008sparse,yoshimura2012reconstruction,mejia2018decoding,yoshimura2017decoding,toda2011reconstruction,umeda2019decoding,hashemi2021unification,qu2022nonnegative,wang2023sparse}, which utilizes the sparsity-promoting prior distribution on model parameter and calculates the maximum a posteriori (MAP) estimation. Thus, the model training and covariate selection can be realized simultaneously.

Another significant problem for neural decoding tasks is that brain recordings are extremely noisy and contain non-Gaussian noises \cite{ball2009signal,liu2016noise}, conflicting with the traditional Gaussian error assumption utilized in sparse Bayesian learning. Consequently, this may cause a significant performance degradation for brain activity decoding. Although preprocessing methods have been developed to denoise the brain recordings, such as independent component analysis (ICA) \cite{jung2000removing}, it is still hard to guarantee that all recorded noise can be removed favorably by preprocessing.

The goal of this paper is to propose a robust implementation of sparse Bayesian learning, so that the covariate selection can be realized with alleviating the negative effect of the recording noise. The motivation of this research is maximum correntropy criterion (MCC), which was proposed in information theoretic learning (ITL) \cite{principe2010information} and has achieved extraordinary robustness in many machine learning scenarios, including regression \cite{liu2007correntropy,li2023partial}, classification \cite{singh2014c,xu2016robust,li2023correntropy}, dimensionality reduction \cite{he2011robust}, etc. This study aims to integrate MCC into the sparse Bayesian learning framework so that sparseness and robustness could be realized simultaneously. This proposal was proved effective for classification tasks by our previous study \cite{li2023correntropy}, and the present paper aims to extend the proposed sparse Bayesian correntropy learning framework into the reconstruction (regression) scene. Main contributions of the present paper are outlined as follows:

\begin{itemize}
\item[1.]
Our preliminary work \cite{li2023correntropy} lacked an interpretation of this proposal, i.e., using MCC in the sparse Bayesian learning. This paper aims to provide an insightful interpretation for this integration, by deriving the error assumption in MCC.
\item[2.]
MCC-based robust likelihood function is utilized with the automatic relevance determination (ARD) prior technique for robust sparse regression with only one hyperparameter to be tuned.
\item[3.]
While the conference paper for the present study \cite{li2023adaptive} has made the theoretical advancements, it lacked a systematic validation of the proposed method in the real-world neural decoding applications, which only used a toy dataset with a noisy output. To fully demonstrate the practical impacts of our method for neural decoding, the present paper aims at a comprehensive evaluation, which includes a synthetic dataset with a noisy input and a real-world muscle activity reconstruction task using cortical current source and EEG, respectively.
\end{itemize}

\section{Method}
\label{sec:method}

\subsection{Problem Setting}
\label{sec:prob}
Regression aims at learning a model from the dataset which can realize a prediction for continuous response variables. One can consider the canonical linear regression model as follows:
\begin{equation}
\label{equ:lr}
t=\mathbf{x}\mathbf{w}+\epsilon
\end{equation}
where $t$ is the model output, $\mathbf{x}=(x_1,x_2,\cdots,x_D)\in \mathbb{R}^{1\times D}$ is a $D$-dimensional covariate, $\mathbf{w}=(w_1,w_2,\cdots,w_D)^T$ $\in \mathbb{R}^{D\times 1}$ is the model parameter, while $\epsilon$ denotes the error term. Usually, one has a finite dataset $\{(\mathbf{x}_n,t_n)\}_{n=1}^N$, with $N$ samples to train the regression model. The dataset can be denoted by $(\mathbf{X},\mathbf{t})$, in which $\mathbf{t}=(t_1,t_2,\cdots,t_N)^T\in \mathbb{R}^{N\times 1}$ contains all the response variables, and $\mathbf{X}\in \mathbb{R}^{N\times D}$ denotes the collection of covariates, where each row is an individual covariate $\mathbf{x}_n$.

\subsection{Conventional Sparse Bayesian Learning with ARD}
\label{sec:sbl}
To construct the relationship between covariate $\mathbf{x}$ and output $t$, Bayesian learning methods first need to make an assumption for the error term $\epsilon$. The most common assumption for $\epsilon$ is to leverage the zero-mean Gaussian distribution with the variance $\sigma^2$. Thus, the probability density function (PDF) of $t$ could be written by $p(t|\mathbf{x})=\mathcal{N}(t|\mathbf{xw},\sigma^2)$, which denotes the Gaussian distribution over $t$ with mean value $\mathbf{xw}$ and variance $\sigma^2$. Then the likelihood is written by assuming the sample independence
\begin{equation}
\label{equ:likelihood}
p(\mathbf{t}|\mathbf{w},\sigma^2)=(2\pi\sigma^2)^{-N/2}\exp\{-\frac{1}{2\sigma^2}\|\mathbf{t}-\mathbf{X}\mathbf{w}\|^2\}
\end{equation}
Maximum likelihood estimation (MLE) could give the solution that maximizes the probability of occurrence for the data under the corresponding error assumption. The Gaussian assumption based MLE is also equivalent to the least-square (LS) criterion and the mean squared error (MSE) loss function.

To acquire a sparse solution, one can further uses a sparsity- promoting prior distribution on the model parameter, and then calculate the MAP estimation. For example, one could assume $\mathbf{w}$ obeys a Laplace distribution, which is equal to $L_1$-regulariz- ation. Here, we employ the automatic relevance determination (ARD), since it can exclude the hyperparameter which controls the sparseness. ARD is a hierarchical prior technique that first assumes zero-mean  Gaussian distribution for each entry of the model parameter with individual variances
\begin{equation}
\label{equ:ard1}
p(\mathbf{w}|\mathbf{a})=\prod_{d=1}^{D}p(w_d|a_d) = \prod_{d=1}^{D}\mathcal{N}(w_d|0,a_d^{-1})
\end{equation}
where $a_d$ is the inverse variance (called relevance parameter) for the distribution of $w_d$, and $\mathbf{a}=(a_1,a_2,\cdots,a_D)$. One can find that $a_d$ regulates the distribution range for corresponding $w_d$. Each $a_d$ is further assigned with the non-informative prior
\begin{equation}
\label{equ:ard2}
p(\mathbf{a})=\prod_{d=1}^{D}p(a_d) =\prod_{d=1}^{D} a_d^{-1}
\end{equation}
This is essentially an improper prior since its integral is infinite and thus can not be normalized \cite{gelman1995bayesian}. The error variance is also assumed with the non-informative prior
\begin{equation}
\label{equ:ard3}
p(\sigma^2) = (\sigma^2)^{-1}
\end{equation}
By defining the likelihood function and the prior distributions, the posterior distribution of all unknown variables is calculated
\begin{equation}
\label{equ:all_post}
p(\mathbf{w},\mathbf{a},\sigma^2|\mathbf{t})=\frac{p(\mathbf{t}|\mathbf{w},\mathbf{a},\sigma^2)p(\mathbf{w},\mathbf{a},\sigma^2)}{p(\mathbf{t})}
\end{equation}
However, because the normalizing integral for the denominator $p(\mathbf{t})=\int p(\mathbf{t}|\mathbf{w},\mathbf{a},\sigma^2)p(\mathbf{w},\mathbf{a},\sigma^2)d\mathbf{w}d\mathbf{a}d\sigma^2$ is intractable, one needs approximation for the posterior distribution to obtain the MAP estimation. This issue can be effectively solved by type-II maximum likelihood \cite{tipping1999relevance,tipping2001sparse}, expectation maximum (EM) \cite{figueiredo2003adaptive}, and variational inference \cite{bishop2000variational}. During the model training, some $a_d$ will become tremendously large, which indicates that the corresponding $w_d$ is tightly distributed around zero. Hence, these dimensions could be removed from the data. In practice, one usually uses a threshold, and prune the features when their corresponding $a_d$ exceeds the threshold. Because this threshold does not make an obvious influence on the sparseness provided it is set large enough (e.g., $10^6$ or $10^8$), ARD does not require the hyper-parameter tuning, thus realizing adaptive sparseness.

\subsection{Maximum Correntropy Criterion}
\label{sec:mcc}
The key concept of MCC is correntropy which was proposed as a generalized correlation function for stochastic process \cite{santamaria2006generalized} and was further extended as a similarity measure between two random variables \cite{liu2007correntropy}. Specifically, given the random variables $a$ and $b$ accompanied with their $N$ observations $\{(a_n,b_n)\}_{n=1}^N$, the correntropy between $a$ and $b$ is defined by
\begin{equation}
\label{equ:correntropy}
V(a,b)\triangleq\mathbb{E}_{p(a,b)}\left[k(a,b)\right]
\end{equation}
which is the expectation of the kernel function $k(\cdot,\cdot)$ between $a$ and $b$ with respect to their joint distribution $p(a,b)$. Usually, the kernel function is practiced by the Gaussian kernel function and the empirical estimation of $V(a,b)$ is obtained by
\begin{equation}
\begin{split}
\label{equ:correntropy_est}
\hat{V}(a,b)&=\frac{1}{N}\sum_{n=1}^{N}{k_h(a_n,b_n)} \; \\
&=\frac{1}{N}\sum_{n=1}^{N}{\exp(-\frac{(a_n-b_n)^2}{2h})} \; \\
\end{split}
\end{equation}
in which $k_h(a,b)\triangleq \exp(-(a-b)^2/2h)$ denotes the Gaussian kernel function with the kernel bandwidth $h>0$.

Correntropy provides an approach for robust data characteri- zation, since the correntropy value is mainly determined along $a=b$ by the Gaussian kernel function $k_h(a,b)$, which exhibits a local property and thus suppresses the negative effect of large deviation that is caused by adverse noise. Correntropy extracts more information from the dataset than conventional 2nd-order statistics since it contains all the even-numbered moments (by Taylor expansion on the kernel function). Thus, one can utilize correntropy as the objective function in the supervised machine learning to maximize the similarity between the prediction and desired target, called maximum correntropy criterion (MCC).

\subsection{Error Assumption in MCC}
\label{sec:mcc_noise}
\begin{figure}[t!]
\centering
\includegraphics[width=0.3\textwidth]{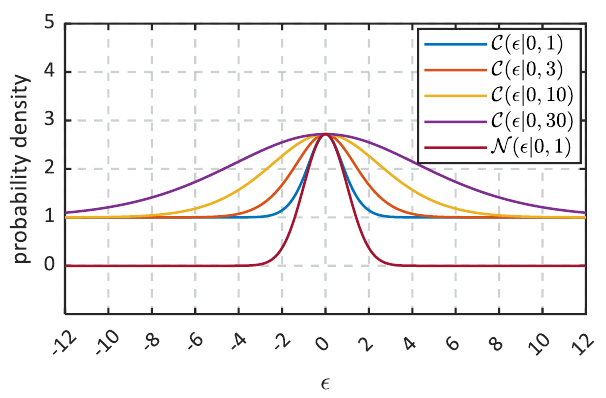}
\caption{$\mathcal{C}(\epsilon|0,h)$ under different $h$ values compared to standard Gaussian distribution $\mathcal{N}(\epsilon|0,1)$. Gaussian distribution is scaled to show the same maximum value as $\mathcal{C}(\epsilon|0,h)$ for comparison.}
\label{fig_mcc_dist}
\end{figure}

To integrate MCC into the sparse Bayesian learning context with theoretical interpretation, we first concentrate on deriving the explicit error assumption inherent in the operation of MCC. MCC-based linear regression, named as maximum correntropy regression (MCR), can be expressed by the following objective function:
\begin{equation}
\begin{split}
\label{equ:mcc_noise_1}
\mathbf{w}&=arg\max _{\mathbf{w}} \frac{1}{N}\sum_{n=1}^{N}{\exp(-\frac{(t_n-\mathbf{x}_n\mathbf{w})^2}{2h})} \; \\
&=arg\max _{\mathbf{w}} \frac{1}{N}\sum_{n=1}^{N}{\exp(-\frac{\epsilon_n^2}{2h})} \; \\
\end{split}
\end{equation}
where $\epsilon_n\triangleq t_n-\mathbf{x}_n\mathbf{w}$ denotes the $n$-th prediction error. Given a fixed dataset $\{(\mathbf{x}_n,t_n)\}_{n=1}^N$, the denominator $N$ is a constant in the objective function and thus could be omitted. As a result, (\ref{equ:mcc_noise_1}) could be transformed to a multiplication form by utilizing an extra exponential function
\begin{equation}
\begin{split}
\label{equ:mcc_noise_2}
\mathbf{w}&=arg\max _{\mathbf{w}} \frac{1}{N}\sum_{n=1}^{N}{\exp(-\frac{\epsilon_n^2}{2h})} \; \\
&=arg\max _{\mathbf{w}}\sum_{n=1}^{N}{\exp(-\frac{\epsilon_n^2}{2h})} \; \\
&=arg\max _{\mathbf{w}}\prod_{n=1}^{N}{\exp\{\exp(-\frac{\epsilon_n^2}{2h})\}} \; \\
\end{split}
\end{equation}
This step is simple but enlightening that transfers the empirical estimation of correntropy to be an accumulative multiplication. It provides a new perspective to interpret MCC as a likelihood function. One can find extraordinarily that (\ref{equ:mcc_noise_2}) can be regarded as a novel MLE if the following error assumption $\mathcal{C}$ is utilized
\begin{equation}
\label{equ:mcc_likelihood}
\mathcal{C}(\epsilon|0,h)\triangleq \exp\{\exp(-\frac{\epsilon^2}{2h})\}
\end{equation}
Here, this novel error assumption $\mathcal{C}$ is named as a correntropy- aware PDF. $\mathcal{C}(\epsilon|0,h)$ denotes the distribution for $\epsilon$ with a zero mean and a shape parameter $h$. Apparently, the integral of this novel PDF is infinite. Further, it is even more improper because when $\epsilon$ becomes arbitrarily large, the probability density would be close to $1$ instead of the normal case $0$. Hence, this proposed PDF $\mathcal{C}$ is called as a deviant error assumption. Fig. \ref{fig_mcc_dist} provides some examples for $\mathcal{C}(\epsilon|0,h)$ under different $h$ values compared to scaled Gaussian distribution (with a same maximum value).

\subsection{Maximum Correntropy Regression with ARD}
\label{sec:mcc_bayes}
By exposing the inherent error assumption in MCR, one can then formally formulate a robust likelihood function motivated by the MCC. Using the error assumption $\mathcal{C}(\epsilon|0,h)$, in the linear regression model, the distribution of $t$ is $p(t|\mathbf{x})=\mathcal{C}(t|\mathbf{xw},h)$. By assuming the independence between each sample, one will obtain the following robust likelihood function
\begin{equation}
\begin{split}
\label{equ:mcc_likelihood_function}
p(\mathbf{t}|\mathbf{w},h)&=\prod_{n=1}^{N}\mathcal{C}(t_n|\mathbf{x}_n\mathbf{w},h) \; \\
&=\prod_{n=1}^{N}\exp\{\exp(-\frac{(t_n-\mathbf{x}_n\mathbf{w})^2}{2h})\}   \; \\
\end{split}
\end{equation}
One can observe that the maximum for this likelihood function is exactly equal to the original objective function of MCC (\ref{equ:mcc_noise_1}).

In what follows, we focus on deriving the integration of the MCC-based likelihood function with ARD technique to realize robust sparse regression from the Bayesian learning viewpoint. Using the robust likelihood (\ref{equ:mcc_likelihood_function}), however, results in a problem that the posterior distribution over $\mathbf{w}$ cannot be obtained analy- tically, since likelihood and prior are not conjugated as before. To address this issue, the variational inference \cite{gelman1995bayesian} is leveraged to approximate the posterior distributions. For simplicity, first, one can consider the shape parameter $h$ (i.e. kernel bandwidth) as a fixed parameter.

Variational inference approximates the true posterior distribution through a surrogate function. To be specific, to estimate and maximize the posterior distribution $p(\mathbf{w},\mathbf{a}|\mathbf{t},h)$, the proxy distribution $q(\mathbf{w},\mathbf{a})$ is introduced. To maximize the similarity between $p(\mathbf{w},\mathbf{a}|\mathbf{t},h)$ and $q(\mathbf{w},\mathbf{a})$, the usual way is minimizing their Kullback-Leibler divergence. Equivalently, one could also minimize the following free energy function \cite{gelman1995bayesian}
\begin{equation}
\begin{split}
\label{equ:mcc_free_energy}
F(q(\mathbf{w},\mathbf{a})) &\triangleq -\mathbb{E}_{q(\mathbf{w},\mathbf{a})}\left[\log\frac{p(\mathbf{w},\mathbf{a},\mathbf{t},h)}{q(\mathbf{w},\mathbf{a})}\right]\; \\
&=-\int q(\mathbf{w},\mathbf{a})\log\frac{p(\mathbf{w},\mathbf{a},\mathbf{t},h)}{q(\mathbf{w},\mathbf{a})}d\mathbf{w}d\mathbf{a} \; \\
\end{split}
\end{equation}
where $p(\mathbf{w},\mathbf{a},\mathbf{t},h)$ denotes the joint distribution. Furthermore, one could apply an extra factorization on the proxy distribution $q(\mathbf{w},\mathbf{a})$=$q_{\mathbf{w}}(\mathbf{w})q_{\mathbf{a}}(\mathbf{a})$. Thus, one has the following free energy
\begin{equation}
\begin{split}
\label{equ:mcc_free_energy_new}
F&(q_{\mathbf{w}}(\mathbf{w})q_{\mathbf{a}}(\mathbf{a})) = \; \\
&-\int q_{\mathbf{w}}(\mathbf{w})q_{\mathbf{a}}(\mathbf{a})\log\frac{p(\mathbf{w},\mathbf{a},\mathbf{t},h)}{q_{\mathbf{w}}(\mathbf{w})q_{\mathbf{a}}(\mathbf{a})}d\mathbf{w}d\mathbf{a} \; \\
\end{split}
\end{equation}
Consequently, the free energy could be minimized with respect to $q_{\mathbf{w}}(\mathbf{w})$ and $q_{\mathbf{a}}(\mathbf{a})$ alternately by
\begin{equation}
\begin{split}
\label{equ:qw_qa_1}
\textit{w-step:}\;&\log q_{\mathbf{w}}(\mathbf{w})=\mathbb{E}_{q_{\mathbf{a}}(\mathbf{a})}\left[\log p(\mathbf{w},\mathbf{a},\mathbf{t},h)\right]+const \; \\
\textit{a-step:}\;&\log q_{\mathbf{a}}(\mathbf{a})=\mathbb{E}_{q_{\mathbf{w}}(\mathbf{w})}\left[\log p(\mathbf{w},\mathbf{a},\mathbf{t},h)\right]+const \; \\
\end{split}
\end{equation}
in which the logarithm of the joint distribution is
\begin{equation}
	\begin{split}
		\label{equ:log_joint}
		&\log p(\mathbf{w},\mathbf{a},\mathbf{t},h)=\log p(\mathbf{t}|\mathbf{w},h)+\log p(\mathbf{w}|\mathbf{a}) +\log p(\mathbf{a}) \; \\
		=&\sum_{n=1}^{N}\exp(-\frac{(t_n-\mathbf{x}_n\mathbf{w})^2}{2h})-\frac{1}{2}\mathbf{w}^T\mathbf{A}\mathbf{w}-\frac{1}{2}\log|\mathbf{A}|+const \; \\
	\end{split}
\end{equation}
where $\mathbf{A}$ is a $D\times D$ diagonal matrix with the diagonal element $(a_1,a_2,\cdots,a_D)$. Then one has
\begin{equation}
\begin{split}
\label{equ:qw_qa_2}
\log q_{\mathbf{w}}(\mathbf{w})&=\sum_{n=1}^{N}\exp(-\frac{(t_n-\mathbf{x}_n\mathbf{w})^2}{2h})-\frac{1}{2}\mathbf{w}^T\mathbb{E}_{q_{\mathbf{a}}(\mathbf{a})}\left[\mathbf{A}\right]\mathbf{w} \; \\
\log q_{\mathbf{a}}(\mathbf{a})&=-\frac{1}{2}\sum_{d=1}^{D}a_d          \mathbb{E}_{q_{\mathbf{w}}(\mathbf{w})}\left[w_d^2\right]-\frac{1}{2}\sum_{d=1}^{D}\log a_d \; \\
\end{split}
\end{equation}
To obtain the expectation $\mathbb{E}_{q_{\mathbf{w}}(\mathbf{w})}\left[w_d^2\right]$, one needs to formulate the distribution for $q_{\mathbf{w}}(\mathbf{w})$, which still cannot be clarified with an analytical form. Therefore, the Laplacian approximation is applied to $\log q_{\mathbf{w}}(\mathbf{w})$ by a quadratic form
\begin{equation}
\label{equ:qw_lap}
\log q_{\mathbf{w}}(\mathbf{w})\approx \log q_{\mathbf{w}}(\mathbf{w}^*)-\frac{1}{2}(\mathbf{w}-\mathbf{w}^*)^T\mathbf{H}(\mathbf{w}^*)(\mathbf{w}-\mathbf{w}^*)
\end{equation}
where $\mathbf{w}^*$ indicates the value that maximizes $\log q_{\mathbf{w}}(\mathbf{w})$, while $\mathbf{H}(\mathbf{w}^*)$ is the negative Hessian matrix of $\log q_{\mathbf{w}}(\mathbf{w})$ at $\mathbf{w}^*$
\begin{equation}
\begin{split}
\label{equ:negative_hessian}
&\mathbf{H}(\mathbf{w})=-\frac{\partial^2\log q_{\mathbf{w}}(\mathbf{w})}{\partial\mathbf{w}\partial\mathbf{w}^T} \; \\
&=-\frac{1}{h}\sum_{n=1}^{N}{\mathbf{x}_n^T\left\{ \exp(-\frac{\epsilon_n^2}{2h})(\frac{\epsilon_n^2}{h}-1)\right\}\mathbf{x}_n}+\mathbb{E}_{q_{\mathbf{a}}(\mathbf{a})}\left[\mathbf{A}\right] \; \\
\end{split}
\end{equation}
where $\epsilon_n$ is the prediction error produced from the current $\mathbf{w}$. As a result, $q_{\mathbf{w}}(\mathbf{w})$ is approximated by a Gaussian distribution $q_{\mathbf{w}}(\mathbf{w})\approx\mathcal{N}(\mathbf{w}|\mathbf{w}^*,\mathbf{H}(\mathbf{w}^*)^{-1})$. The expectation $\mathbb{E}_{q_{\mathbf{w}}(\mathbf{w})}\left[w_d^2\right]$ is calculated by
\begin{equation}
\mathbb{E}_{q_{\mathbf{w}}(\mathbf{w})}\left[w_d^2\right]=w_d^{*2}+s_d^2
\end{equation}
where $s_d^2$ is the $d$-th diagonal element in $\mathbf{H}(\mathbf{w}^*)^{-1}$. To obtain the maximum solution $\mathbf{w}^*$ for $\log q_{\mathbf{w}}(\mathbf{w})$, one can observe that this is in essence an $L_2$-regularized MCR and $\mathbb{E}_{q_{\mathbf{a}}(\mathbf{a})}\left[\mathbf{A}\right]$ plays the role of $L_2$-regularization weight. The following fixed-point method provides an effective way to optimize $\mathbf{w}$ with very fast convergence \cite{chen2015convergence}
\begin{equation}
\label{equ:w_fp}
\mathbf{w}=(\mathbf{X}^T\mathbf{\Psi}\mathbf{X}+\mathbb{E}_{q_{\mathbf{a}}(\mathbf{a})}\left[\mathbf{A}\right])^{-1}\mathbf{X}^T\mathbf{\Psi}\mathbf{t}
\end{equation}
in which $\mathbf{\Psi}$ denotes a $N\times N$ diagonal matrix, and its diagonal element is $\Psi_{nn}=\exp(-\epsilon_n^2/2h)$.

On the other hand, $q_{\mathbf{a}}(\mathbf{a})$ is analytically equal to the Gamma distribution as follows by applying an exponential function on $\log q_{\mathbf{a}}(\mathbf{a})$ in (\ref{equ:qw_qa_2})
\begin{equation}
\label{equ:a_gamma}
q_{\mathbf{a}}(\mathbf{a})=\prod_{d=1}^{D}{q_{a_d}(a_d)}=\prod_{d=1}^{D}{\varGamma(a_d|a_d^*,\frac{1}{2})}
\end{equation}
where $\varGamma(a_d|a_d^*,\frac{1}{2})$ is the Gamma distribution over $a_d$ with the expectation $a_d^*$ and the degree of freedom $\frac{1}{2}$. The expectation $a_d^*$ is computed by
\begin{equation}
\label{equ:slr_a_star}
a_d^*=\frac{1}{w_d^{*2}+s_d^2}
\end{equation}
which can also be updated by the following equation for faster convergence \cite{wipf2007new,tipping2001sparse}
\begin{equation}
\label{equ:a_fast}
a_d^*=\frac{1-a_d^*s_d^2}{w_d^{*2}}
\end{equation}
The updated $a_d$ is substituted into $\mathbb{E}_{q_{\mathbf{a}}(\mathbf{a})}\left[\mathbf{A}\right]$ for \textit{w-step} in the next iteration. By optimizing \textit{w-step} and \textit{a-step} alternately, the free energy will be minimized, while the MAP estimations for $\mathbf{w}$ and $\mathbf{a}$ will be also obtained, respectively. Algorithm \ref{cslr} gives a summary for the proposed method that integrates MCR with ARD in a Bayesian framework.

\begin{algorithm}[h]
	\caption{\textit{MCR-ARD for Robust Sparse Regression}}
	\label{cslr}
	\begin{algorithmic}[1]
		\State \textbf{input}:
		
		training samples $\{(\mathbf{x}_n,t_n)\}_{n=1}^N$;
		
		kernel bandwidth $h$;
		
		threshold for covariate pruning $a_{\max}$;
		
		\State \textbf{initialize}:
		
		model parameter $w_d$ ($d=1,\cdots,D$);
		
		relevance parameter $a_d=1$ ($d=1,\cdots,D$);
		
		\State \textbf{output}:
		
		model parameter $w_d$ ($d=1,\cdots,D$)
		\Repeat 
		\State \textit{w-step}: update $\mathbf{w}$ according to (\ref{equ:w_fp});
		\State \textit{a-step}: update $\mathbf{a}$ according to (\ref{equ:a_fast});
		\If{$a_d\geqslant a_{\max}$}
		\State fix the corresponding $w_d$ to be zero and prune this dimension in the following optimization
		\EndIf
		\Until the parameter change is small enough or the  iteration number is larger than a predetermined limit
	\end{algorithmic}
\end{algorithm}

\subsection{Performance Evaluation}
\label{sec:eval}

Three performance evaluation scenarios were utilized in this study to fully validate our proposed MCR-ARD algorithm. The three evaluation datasets include a synthetic toy with corrupted covariate, a real-world muscle activity reconstruction task from estimated cortical current source, and the same muscle activity reconstruction task utilizing the EEG recording. The latter two were proposed in our previous study \cite{yoshimura2012reconstruction}, which employed the conventional LSR-ARD to achieve the brain activity decoding.

For performance evaluation, MCR-ARD was compared with its baseline, the conventional sparse Bayesian regression which employs the Gaussian likelihood (least-square criterion), called LSR-ARD, as introduced in Section \ref{sec:sbl}. Both two algorithms used an identical pruning threshold $a_{\max}$ of $10^6$. The maximal iteration number was set to be $500$ for both algorithms. Section \ref{sec:dis} provides a discussion about the relationship between MCR-ARD and other existing methods.

With regard to the performance indicators for evaluation, the two most commonly used ones in regression were utilized, i.e., correlation coefficient$=Cov(\mathbf{\hat{t}},\mathbf{t})/\sqrt{Var(\mathbf{\hat{t}})Var(\mathbf{t})}$, and root mean squared error$=\sqrt{\lVert \mathbf{\hat{t}} - \mathbf{t} \rVert^2/N}$, for which $\mathbf{\hat{t}}$ indicates the model prediction for $\mathbf{t}$, while $Cov(\cdot,\cdot)$ and $Var(\cdot)$ denote the covariance and variance, respectively.

\subsubsection{Synthetic Dataset}~
\label{sec:toy_des}

A demo concerning this synthetic dataset is provided\footnote{\href{https://sites.google.com/view/liyuanhao/code}{\textcolor{IEEEBlue}{\textit{https://sites.google.com/view/liyuanhao/code}}}} which could be utilized to easily reproduce the corresponding results. To evaluate the robustness and the feature selection capability, a noisy and high-dimensional synthetic dataset was considered. Specifically, 300 independent and identically distributed (i.i.d.) training samples with 300 i.i.d. testing samples were generated stochastically using the 500-dimensional standard multivariate Gaussian distribution (i.e., the covariance matrix is an identity matrix). The optimal solution was designed by a sparse format
\begin{equation}
	\label{equ:true_solution}
	\mathbf{w}^*=(\overset{500\; components}{\overbrace{w^*_1,\cdots,w^*_{30},\underset{470\; zeros}{\underbrace{0,0,0,0,\cdots,0}}}})^T
\end{equation}
which means only the first $30$ features are relevant to the task. The non-zero $30$ elements in $\mathbf{w}^*$ was generated by the standard Gaussian distribution. The target $\mathbf{t}$ was obtained as the product between the covariates and the sparse solution $\mathbf{w}^*$. As a result, we trained the regression models on 300 training samples with 500 features, i.e., high-dimensional problem, and validated the models on the remaining 300 testing samples. For MCR-ARD, the kernel bandwidth $h$ was selected by 5-fold cross validation from $30$ values ranging from $1.0$ to $1000$ with equal differences on a logarithmic scale. Specifically, 300 training samples were divided into 5 sets. Each candidate value for kernel bandwidth was utilized for model training on 4 out of the 5 sets, and then evaluated on the other one, until each set is used for validation. The optimal kernel bandwidth was selected which realized the highest average correlation coefficient on 5 validation sets, and then used to train the model utilizing all training samples. Thus the testing data was not used in the kernel bandwidth selection.

Regarding the contamination for this toy dataset, we mainly considered the corrupted covariate matrix, because noisy brain recording is employed as input variable in brain decoding task. The conference paper \cite{li2023adaptive} reported a simulation for corrupted target variable. To realize a realistic corruption on the covariate matrix $\mathbf{X}$, the arbitrary corruption model \cite{chen2013robust} was used, which means that any arbitrary elements in the covariate matrix could be contaminated randomly. To generate noises that are adverse enough on model training, the Laplace distribution was utilized for noise synthesis because the Laplace distribution is a heavy-tailed distribution. Hence, it is more likely to generate adverse outliers for data corruption. To be specific, a certain proportion (from $0$ to $1.0$ with a step $0.1$) of the components for covariate matrix were randomly selected. Then, their values were added by a Laplace-distribution-induced noise. A zero-mean Laplace distribution with the following scale parameters was leveraged, respectively: $0.2$, $0.3$, $0.5$, $0.7$, $1.0$, $1.2$, and $1.5$. A larger scale parameter is easier to induce the noise of large amplitude, thus being more detrimental to the dataset. On the other hand, only the training data was corrupted while the testing data remained unchanged, which is a typical strategy to assess the robustness for different algorithms \cite{zhu2004class}. To obtain generalized results, the above settings were executed for 100 Monte-Carlo repetitions.

\subsubsection{Muscle Activity Reconstruction by Current Source}~
\label{sec:eval2}

\begin{table*}[t]
	\centering
	\caption{Summary of session information with the numbers of estimated sources for the muscle activity reconstruction dataset.}
	\label{session_info}
	\resizebox{1\textwidth}{!}{
		\renewcommand{\arraystretch}{1.2}
		\setlength{\tabcolsep}{3.2pt}
		\begin{tabular}{@{}ccccccccccccccccccccccccc@{}} 
			\toprule
			\hline
			\multicolumn{1}{c|}{\multirow{1}{*}{Task}} & \multicolumn{14}{c|}{\textit{\textbf{Flexion}}} & \multicolumn{10}{c}{\textit{\textbf{Extension}}} \\
			\hline
			\multicolumn{1}{c|}{\multirow{1}{*}{Session}} & \textit{Sess.1} & \textit{Sess.2} & \textit{Sess.3} & \textit{Sess.4} & \textit{Sess.5} & \textit{Sess.6} & \textit{Sess.7} & \textit{Sess.8} & \textit{Sess.9} & \textit{Sess.10} & \textit{Sess.11} & \textit{Sess.12} & \textit{Sess.13} & \multicolumn{1}{c|}{\textit{Sess.14}} & \textit{Sess.1} & \textit{Sess.2} & \textit{Sess.3} & \textit{Sess.4} & \textit{Sess.5} & \textit{Sess.6} & \textit{Sess.7} & \textit{Sess.8} & \textit{Sess.9} & \textit{Sess.10} \\
			\hline
			\multicolumn{1}{c|}{\multirow{1}{*}{Subject}} & \multicolumn{3}{c|}{\textit{Subj.1}} & \multicolumn{4}{c|}{\textit{Subj.2}} & \multicolumn{3}{c|}{\textit{Subj.3}} & \multicolumn{4}{c|}{\textit{Subj.4}} & \multicolumn{1}{c|}{\textit{Subj.1}} & \multicolumn{4}{c|}{\textit{Subj.2}} & \multicolumn{1}{c|}{\textit{Subj.3}} & \multicolumn{4}{c}{\textit{Subj.4}} \\
			\hline
			\multicolumn{1}{c|}{\multirow{1}{*}{Activity}} & \textit{FH} & \textit{FL} & \multicolumn{1}{c|}{\textit{FL}} & \textit{FH} & \textit{FL} & \textit{FH} & \multicolumn{1}{c|}{\textit{FL}} & \textit{FH} & \textit{FL} & \multicolumn{1}{c|}{\textit{FL}} & \textit{FH} & \textit{FL} & \textit{FH} & \multicolumn{1}{c|}{\textit{FL}} & \multicolumn{1}{c|}{\textit{EL}} & \textit{EH} & \textit{EL} & \textit{EH} & \multicolumn{1}{c|}{\textit{EL}} & \multicolumn{1}{c|}{\textit{EL}} & \textit{EH} & \textit{EL} & \textit{EH} & \textit{EL} \\
			\hline
			\multicolumn{1}{c|}{\multirow{1}{*}{$\#$ Sources}} & 63 & 63 & \multicolumn{1}{c|}{59} & 77 & 77 & 130 & \multicolumn{1}{c|}{79} & 137 & 137 & \multicolumn{1}{c|}{117} & 211 & 211 & 199 & \multicolumn{1}{c|}{159} & \multicolumn{1}{c|}{59} & 127 & 127 & 130 & \multicolumn{1}{c|}{79} & \multicolumn{1}{c|}{117} & 112 & 112 & 199 & 159 \\
			\hline
			\bottomrule
			\multicolumn{25}{r}{\multirow{1}{*}{*\textit{FH}=\textit{FlexHigh}, \textit{FL}=\textit{FlexLow}, \textit{EH}=\textit{ExtHigh}, \textit{EL}=\textit{ExtLow}}}\\
	\end{tabular}}
\end{table*}

\begin{figure*}[t!]
	\centering
	\includegraphics[width=1\textwidth]{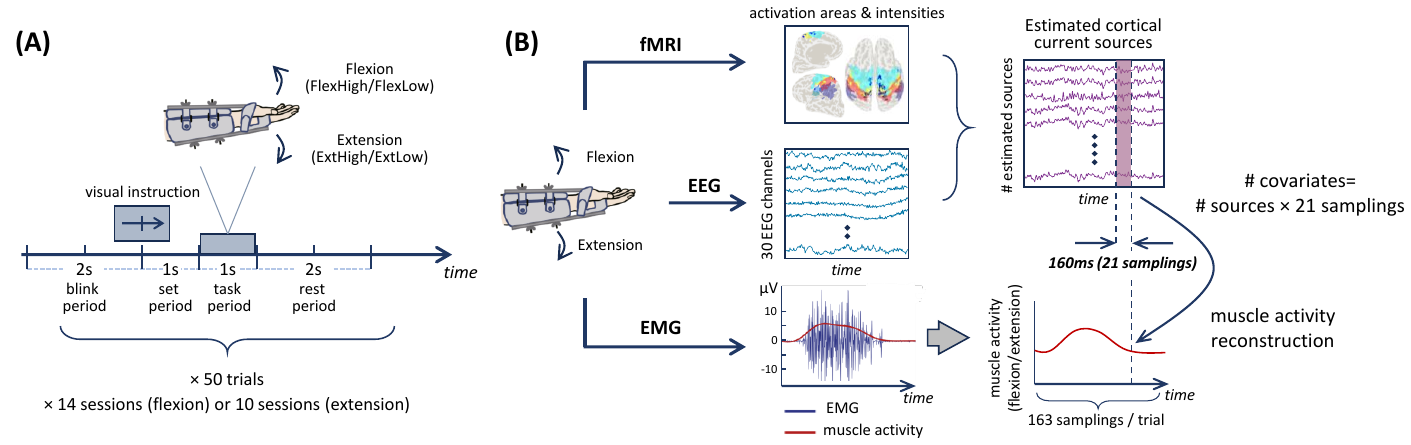}
	\caption{Schematic overview for the real-world muscle activity reconstruction task:(A) Trial diagram of the EEG-EMG experiment (B) Decoding paradigm of muscle activity reconstruction utilizing cortical current sources estimated from EEG and fMRI data.}
	\label{fig_data}
\end{figure*}

Five right-handed subjects participated in our previous study \cite{yoshimura2012reconstruction}. The experimental protocol was approved by a local ethics committee of the National Center of Neurology and Psychiatry, and all subjects gave written informed consent. Due to the data storage issue, only the data of four subjects was utilized in this study. Right wrist of each subject was immobilized with plastic casts that allowed the subject to perform isometric contraction tasks. The experiments were comprised of two parts, one using EEG and the other using fMRI. In addition, EMG signals were recorded in both two parts. During the EEG-EMG experiment, according to the visual instructions, each subject was requested to perform flexion and extension using high force or low force, thus resulting in the following four different combinations, i.e., 1) flexion with high force (FlexHigh), 2) flexion with low force (FlexLow), 3) extension with high force (ExtHigh), and finally 4) extension with low force (ExtLow). The muscle activity was recorded at the sampling rate of 5000 Hz by utilizing two pairs of 2 cm-spaced Ag/AgCl EMG electrodes attaching over right flexor carpi radialis and right extensor carpi radialis brevis, the major muscles for wrist flexion and extension. Simultaneously, a 30-channel EEG system positioned according to the extended international 10–20 system was utilized to record brain activity at the sampling rate of 5000 Hz. Two amplifiers for EMG and EEG recordings were both set to the resolution of 0.5 $\mu V$ and the range of $\pm$16 $mV$, respectively. Every trial for EEG-EMG experiment started with a 2 s blink period. Eye blink was only allowed in this period to avoid corruptions on EEG recordings. Then, the subjects received the visual instruction for 1 s, while the wrist kept still, called ``set'' period. In the next 1 s, subjects carried out the instructed action. Finally, the current trial ended by a 2 s rest, and the subsequent trial would start, as illustrated in Fig. \ref{fig_data}(A). Each EEG-EMG session contained 50 trials with a same muscle activity. The session information is summarized in Table \ref{session_info} for each subject with the activity category. Moreover, each subject participated in an fMRI experiment to identify the brain activation area and intensity of high spatial resolution for the cortical current source estimations. A 3 Tesla MRI scanner was used for the fMRI experiment, which consisted of 7 blocks for each subject. During each block, the subject was requested to execute four muscle activities in a pseudo-randomized order.

For data preprocessing, the muscle activities were calculated from the EMG signals utilizing the method \cite{koike1995estimation} for flexion and extension, which were further downsampled to 125 Hz for the following analysis. The last 0.3 s for the set period and all the task period were used for muscle activity reconstruction. Each trial had 163 samplings for the muscle activity as a result (1.3 s $\times$ 125 Hz). The EEG recordings were first bandpass-filtered from 0.5 Hz to 40 Hz and then downsampled to 250 Hz. fMRI data was analyzed to create the area information that contained the regions of statistically significant voxels and identified the source estimation scope. Accordingly, the number of estimated current sources could be different. To estimate cortical current source, a hierarchical Bayesian method \cite{sato2004hierarchical,yoshioka2008evaluation} was utilized. The prior obtained from fMRI was used for the corresponding subject, i.e. individual analysis. Then, the estimated source was downsampled to 125 Hz which would be used as the covariate for the decoding. Table \ref{session_info} lists the number of estimated sources.

To reconstruct the muscle activity from the brain recordings, a similar decoding paradigm as in our previous study \cite{yoshimura2012reconstruction} was utilized. First, an intra-session decoding was considered, where each session was used as a sub-dataset for both model training and testing. For intra-session decoding, only the corresponding muscle activities were reconstructed in performance validation, e.g., only the flexion was predicted for FlexHigh and FlexLow sessions. The target activity was normalized to a range of $[0,1]$ for each session. Afterward each session was randomly divided into 5 batches, each containing 10 trials. The regression model was trained on 4 batches and then tested on the remaining one, until each batch was used for the testing. To predict the muscle activity for each time point, the cortical current sources during the previous 160 ms (21 samplings) were utilized as covariate, as shown in Fig. \ref{fig_data}(B). To select a proper bandwidth for MCR-ARD, 35 random trials were selected from the training batches to train the model, with 30 different $h$ values ranging from $1.0$ to $1000$, as mentioned in Section \ref{sec:toy_des}. The optimal value was selected by validating on the remaining 5 trials for the training batches. Eventually, all 40 training trials were utilized to train MCR-ARD by the optimal kernel bandwidth. Thus, the tuning did not leverage the testing batch. As a result, for each session, five reconstruction models were trained by each algorithm, and 50 testing results were obtained by the 10 testing trials in each loop. Such a procedure was repeated 10 times to acquire more generalized results. Hence, each session resulted in 500 testing results for each algorithm.

\begin{figure*}[t!]
	\centering
	\includegraphics[width=0.92\textwidth]{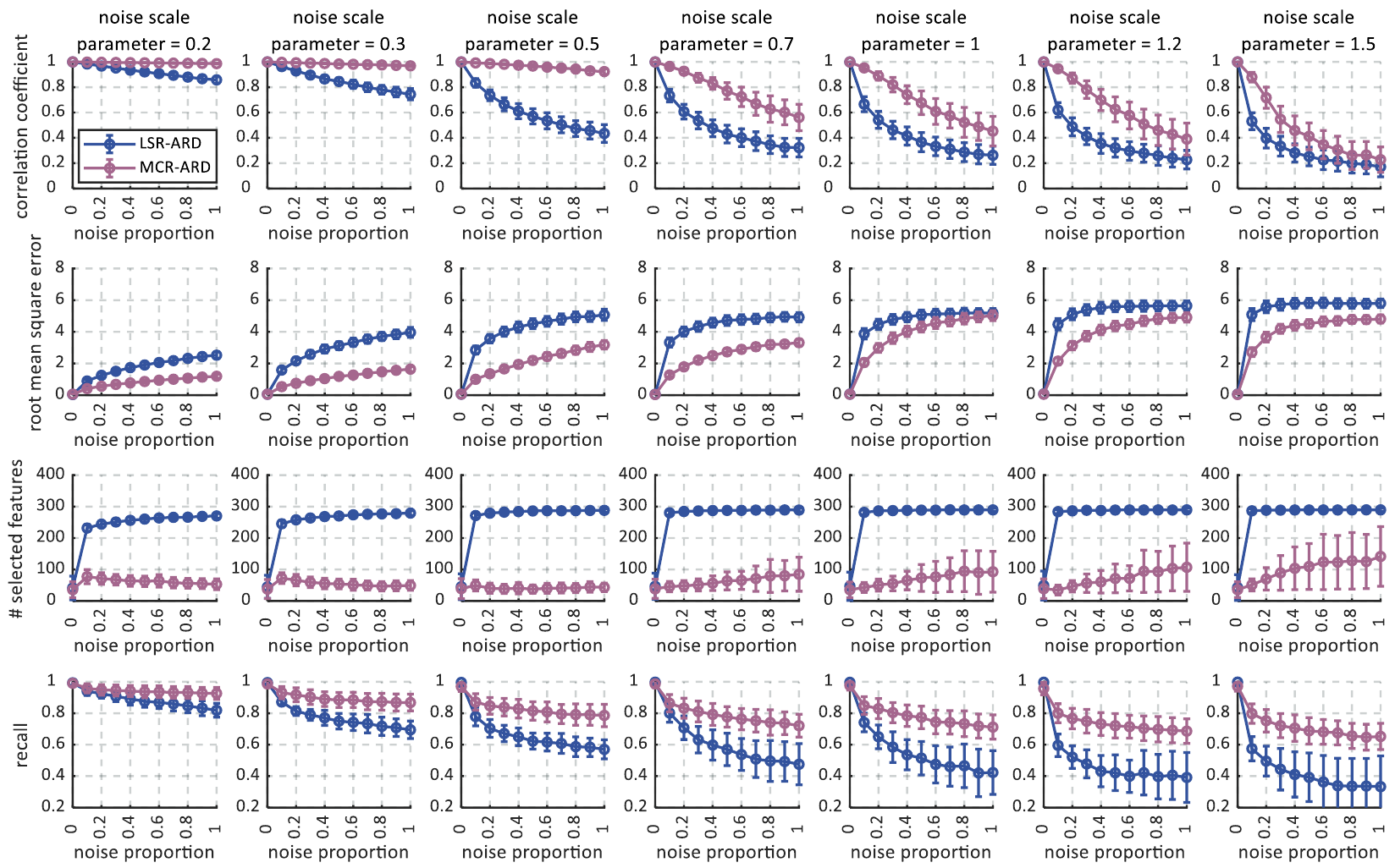}
	\caption{Correlation coefficient, root mean squared error, number of selected features, and feature selection recall for the synthetic dataset by LSR-ARD and MCR-ARD, respectively. The results are averaged by $100$ Monte-Carlo repetitions. Error bar presents the standard deviation for each result.}
	\label{fig_toy}
\end{figure*}

\begin{figure}[t!]
	\centering
	\includegraphics[width=0.36\textwidth]{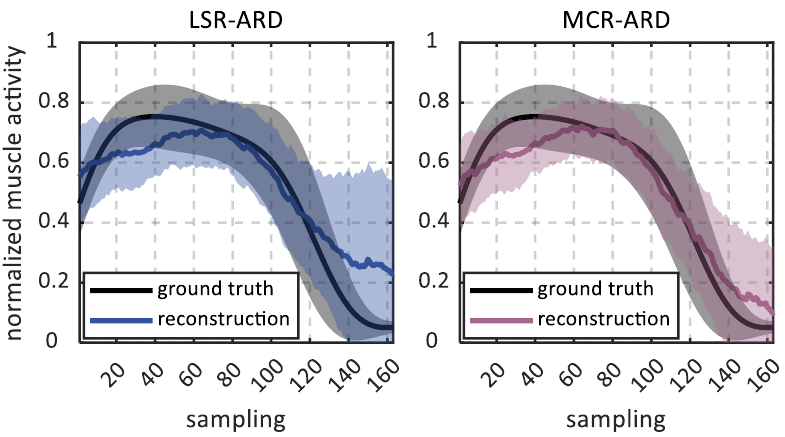}
	\caption{Muscle activity of ground truth and reconstruction from LSR-ARD and MCR-ARD, respectively, for Sess.14 of flexion activity reconstruction. The ground truth illustrates the average of 50 trials in this session, and the reconstructions are averaged across 500 testing results. The shade area denotes the standard deviation.}
	\label{fig_muscle1}
\end{figure}

Furthermore, for a more comprehensive decoding paradigm, inter-session decoding was considered to identify the decoding of different activities (flexion / extension) and different forces (high / low). Specifically, every two sessions of a same number of estimated sources were combined to train and test a muscle activity reconstruction model with a basically same framework as in the intra-session decoding. Table \ref{inter_session_info} lists the combinations of inter-session decoding. Notably after the training and testing batches were divided in each session, they were utilized jointly for each combination, that each had 80 training and 20 testing trials. For the decoding of different forces (Combos.1, 3, 4, 7, 9, 10), only the activated muscle activities were analyzed. For the decoding of two activities (Combos.2, 5, 6, 8, 11, 12), all muscle activities were utilized for prediction. Muscle activities were also normalized into $[0,1]$. By 5-fold cross validation and 10 repetitions, each combination had 1000 testing results. The kernel bandwidth of MCR-ARD was selected in a similar way. Each candidate value was trained on 70 training trials and then validated on the other 10 training trials, which did not involve the testing trials.

\begin{table}[t]
	\centering
	\caption{Combination details of the inter-session framework.}
	\label{inter_session_info}
	\resizebox{0.44\textwidth}{!}{
		\renewcommand{\arraystretch}{1.2}
		\setlength{\tabcolsep}{3.2pt}
		\begin{tabular}{@{}cccc@{}} 
			\toprule
			\hline
			\multicolumn{1}{c|}{\multirow{1}{*}{Combo. Index}} & \multicolumn{1}{c|}{Subject} & \multicolumn{1}{c|}{Task} & Combined Sessions\\
			\hline
			\multicolumn{1}{c|}{\multirow{1}{*}{\textit{Combo.1}}} & \multicolumn{1}{c|}{\multirow{2}{*}{\textit{Subj.1}}} & \multicolumn{1}{c|}{\textit{FH vs. FL}} & \textit{Flexion Sess.1 + Flexion Sess.2}\\
			\multicolumn{1}{c|}{\multirow{1}{*}{\textit{Combo.2}}} & \multicolumn{1}{c|}{\multirow{2}{*}{}} & \multicolumn{1}{c|}{\textit{FL vs. EL}} & \textit{Flexion Sess.3 + Extension Sess.1}\\
			\hline
			\multicolumn{1}{c|}{\multirow{1}{*}{\textit{Combo.3}}} & \multicolumn{1}{c|}{\multirow{4}{*}{\textit{Subj.2}}} & \multicolumn{1}{c|}{\textit{FH vs. FL}} & \textit{Flexion Sess.4 + Flexion Sess.5}\\
			\multicolumn{1}{c|}{\multirow{1}{*}{\textit{Combo.4}}} & \multicolumn{1}{c|}{\multirow{4}{*}{}} & \multicolumn{1}{c|}{\textit{EH vs. EL}} & \textit{Extension Sess.2 + Extension Sess.3}\\
			\multicolumn{1}{c|}{\multirow{1}{*}{\textit{Combo.5}}} & \multicolumn{1}{c|}{\multirow{4}{*}{}} & \multicolumn{1}{c|}{\textit{FH vs. EH}} & \textit{Flexion Sess.6 + Extension Sess.4}\\
			\multicolumn{1}{c|}{\multirow{1}{*}{\textit{Combo.6}}} & \multicolumn{1}{c|}{\multirow{4}{*}{}} & \multicolumn{1}{c|}{\textit{FL vs. EL}} & \textit{Flexion Sess.7 + Extension Sess.5}\\
			\hline
			\multicolumn{1}{c|}{\multirow{1}{*}{\textit{Combo.7}}} & \multicolumn{1}{c|}{\multirow{2}{*}{\textit{Subj.3}}} & \multicolumn{1}{c|}{\textit{FH vs. FL}} & \textit{Flexion Sess.8 + Flexion Sess.9}\\
			\multicolumn{1}{c|}{\multirow{1}{*}{\textit{Combo.8}}} & \multicolumn{1}{c|}{\multirow{2}{*}{}} & \multicolumn{1}{c|}{\textit{FL vs. EL}} & \textit{Flexion Sess.10 + Extension Sess.6}\\
			\hline
			\multicolumn{1}{c|}{\multirow{1}{*}{\textit{Combo.9}}} & \multicolumn{1}{c|}{\multirow{4}{*}{\textit{Subj.4}}} & \multicolumn{1}{c|}{\textit{FH vs. FL}} & \textit{Flexion Sess.11 + Flexion Sess.12}\\
			\multicolumn{1}{c|}{\multirow{1}{*}{\textit{Combo.10}}} & \multicolumn{1}{c|}{\multirow{4}{*}{}} & \multicolumn{1}{c|}{\textit{EH vs. EL}} & \textit{Extension Sess.7 + Extension Sess.8}\\
			\multicolumn{1}{c|}{\multirow{1}{*}{\textit{Combo.11}}} & \multicolumn{1}{c|}{\multirow{4}{*}{}} & \multicolumn{1}{c|}{\textit{FH vs. EH}} & \textit{Flexion Sess.13 + Extension Sess.9}\\
			\multicolumn{1}{c|}{\multirow{1}{*}{\textit{Combo.12}}} & \multicolumn{1}{c|}{\multirow{4}{*}{}} & \multicolumn{1}{c|}{\textit{FL vs. EL}} & \textit{Flexion Sess.14 + Extension Sess.10}\\
			\hline
			\bottomrule
			\multicolumn{4}{r}{\multirow{1}{*}{*\textit{FH}=\textit{FlexHigh}, \textit{FL}=\textit{FlexLow}, \textit{EH}=\textit{ExtHigh}, \textit{EL}=\textit{ExtLow}}}\\
	\end{tabular}}
\end{table}

\subsubsection{Muscle Activity Reconstruction by EEG}~
\label{sec:eval3}

Furthermore, we evaluated the decoding performance for the above-mentioned muscle activity regression task, only utilizing the EEG signal in the decoding by LSR-ARD and MCR-ARD. Specifically, the fMRI recording and the cortical current source estimation procedure were excluded from the above-mentioned decoding paradigm. The pre-processed EEG recordings, which were also downsampled to 125 Hz similarly, were directly used as the covariates to predict the muscle activities, with the same intra-session decoding procedures, as are mentioned in Section \ref{sec:eval2}.

\section{Results}
\label{sec:exp}

\subsection{Synthetic Dataset}
\label{sec:toy}

Fig. \ref{fig_toy} illustrates the regression performance of the synthetic dataset obtained from LSR-ARD and MCR-ARD, respectively, averaged by $100$ Monte-Carlo repetitions. The top row presents the correlation coefficient, while the second row shows the root mean squared error. One perceives that, the conventional LSR-ARD was influenced remarkably by the manual contamination. In contrast, MCR-ARD suppressed  the adverse impact, leading to obviously higher correlation coefficient and lower root mean squared error, indicating a superior prediction than LSR-ARD.

Moreover, the feature selection capability was also evaluated for the synthetic toy. The third row in Fig. \ref{fig_toy} shows the numbers of selected features for different corruption conditions. Further, to quantitatively examine the feature selection result, the recall was calculated for the $30$ relevant dimensions, which indicates what percentage of these relevant features were selected by the sparse algorithm. The bottom row in Fig. \ref{fig_toy} presents the feature selection recall for each algorithm. One can observe that LSR-ARD and MCR-ARD selected a similar number of covariates, and realized high enough recall when the noise proportion was $0$. However, even a small noise proportion, such as $0.1$, would result in an obvious impact on the number of selected features for LSR-ARD, whereas MCR-ARD could realize a much more stable number for selected features. Quantitatively, MCR-ARD achieved a higher feature selection recall than LSR-ARD under artificial corruption, suggesting better feature selection quality.

\subsection{Muscle Activity Reconstruction by Current Source}
\label{sec:cs}

\begin{figure*}[t!]
\centering
\includegraphics[width=0.92\textwidth]{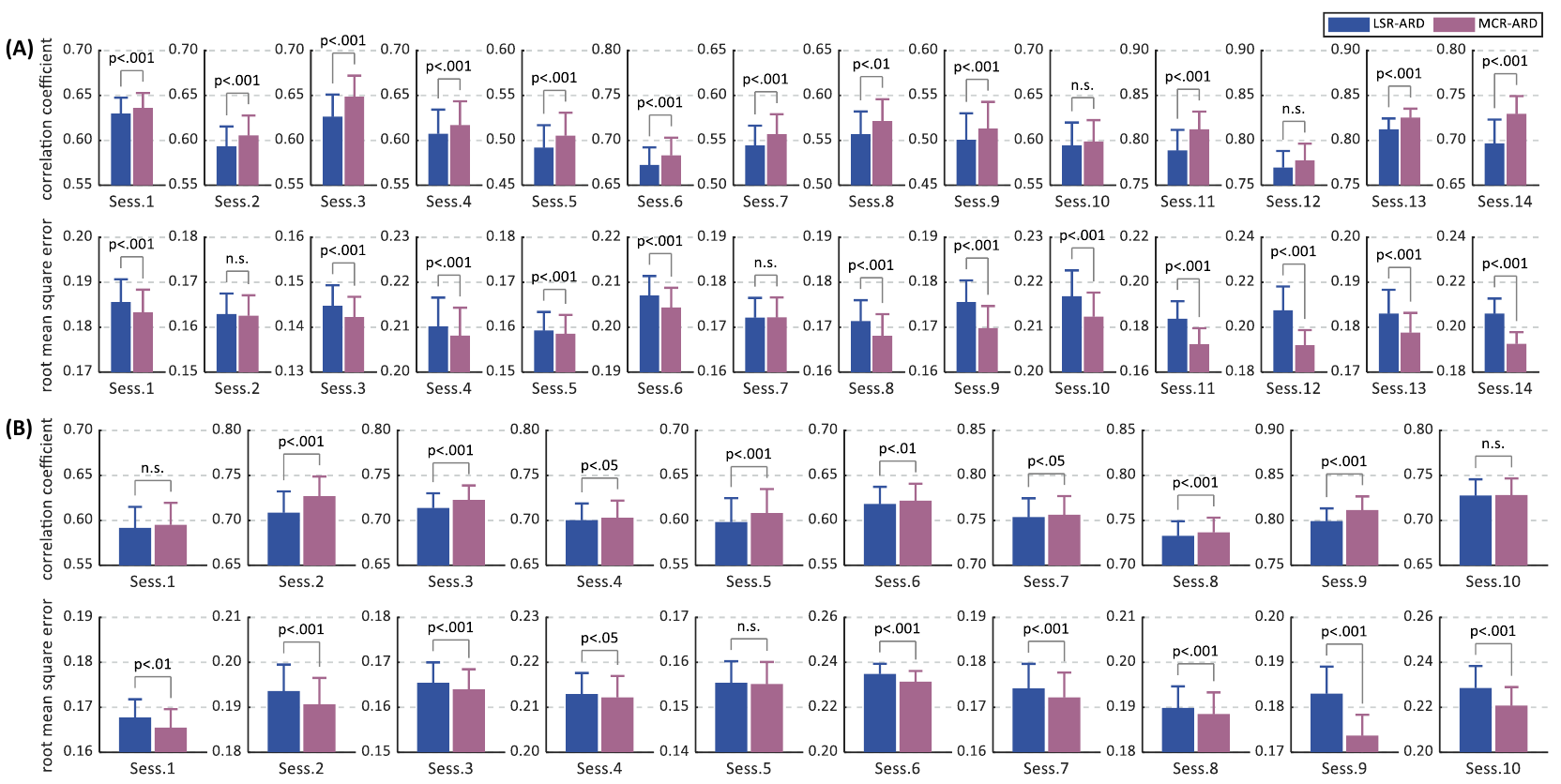}
\caption{Correlation coefficient and root mean squared error for the real-world muscle activity reconstruction dataset by employing the estimated cortical current source signals as the covariates:(A) flexion sessions (B) extension sessions. The results are averaged by repeating the five-fold cross validation for ten repetitions. The error bars denote the $95\%$ confidence intervals for each result. Paired $t$-test was conducted to examine the statistically significant difference. ``n.s.'' indicates no significant difference (\textit{p}$>$.05).}
\label{fig_session}
\end{figure*}

Fig. \ref{fig_muscle1} shows an example for intra-session decoding, Sess.14 for flexion activity reconstruction. Qualitatively, one perceives that the reconstruction from MCR-ARD is closer to the ground truth of the normalized muscle activity with a smaller standard deviation, in particular for the samplings from $\#$120 to $\#$163. For the quantitative performance indicators for all 24 sessions, the reconstruction result of intra-session decoding is illustrated in Fig. \ref{fig_session}, with paired \textit{t}-test to examine the statistical difference between LSR-ARD and MCR-ARD. Out of 14 flexion sessions the proposed MCR-ARD showed higher correlation coefficient and lower root mean squared error than the conventional LSR-ARD with statistically significant difference in 12 sessions. In the 10 extension sessions MCR-ARD outperformed LSR-ARD with significant difference for 8 and 9 sessions, for correlation coefficient and root mean squared error, respectively. Table \ref{table_intra} lists the average decoding performance for each type of muscle activity for each subject, where the superior results are marked in bold font. One observes that MCR-ARD outperformed LSR-ARD regarding all muscle activities on average for all subjects.

\begin{figure*}[t]
\centering
\includegraphics[width=0.96\textwidth]{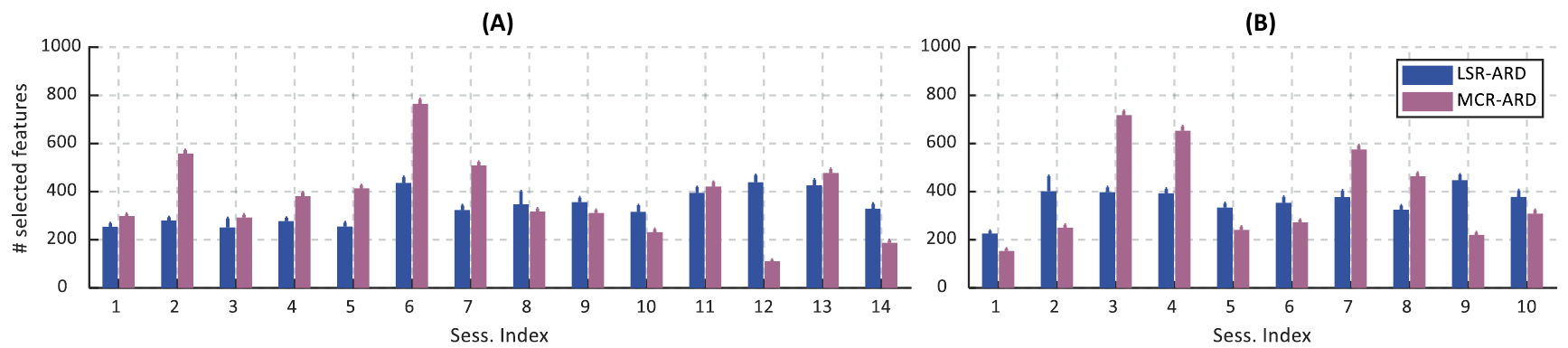}
\caption{Number of selected features averaged across the trained 50 reconstruction models for each session: (A) flexion sessions (B) extension sessions. The error bars indicate the standard deviations for each corresponding result. Note that the total number of features would be different across sessions because the number of estimated current sources varies, as summarized in Table-I.}
\label{figs1}
\end{figure*}

\begin{figure*}[t]
\centering
\includegraphics[width=0.76\textwidth]{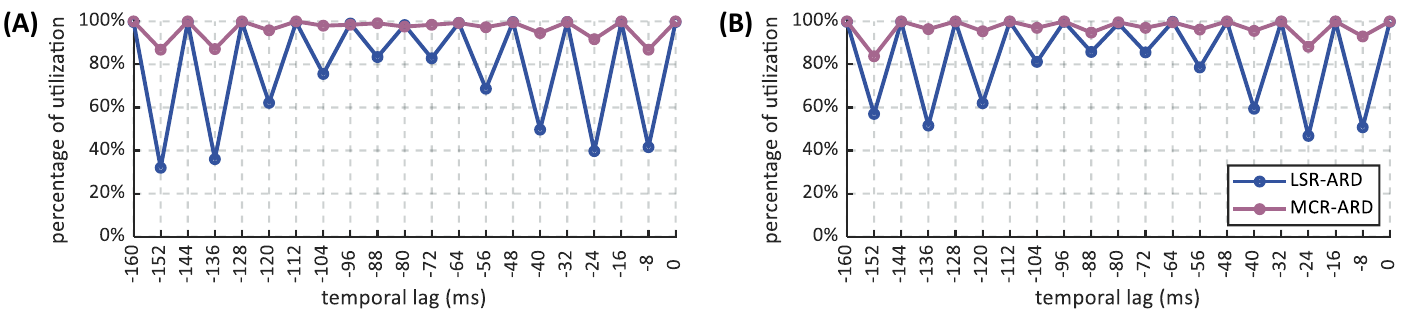}
\caption{Percentage of utilization for each temporal lag in the trained 50 reconstruction models averaged across all sessions:(A) flexion sessions (B) extension sessions.}
\label{figs2}
\end{figure*}

\begin{figure*}[t]
\centering
\includegraphics[width=0.96\textwidth]{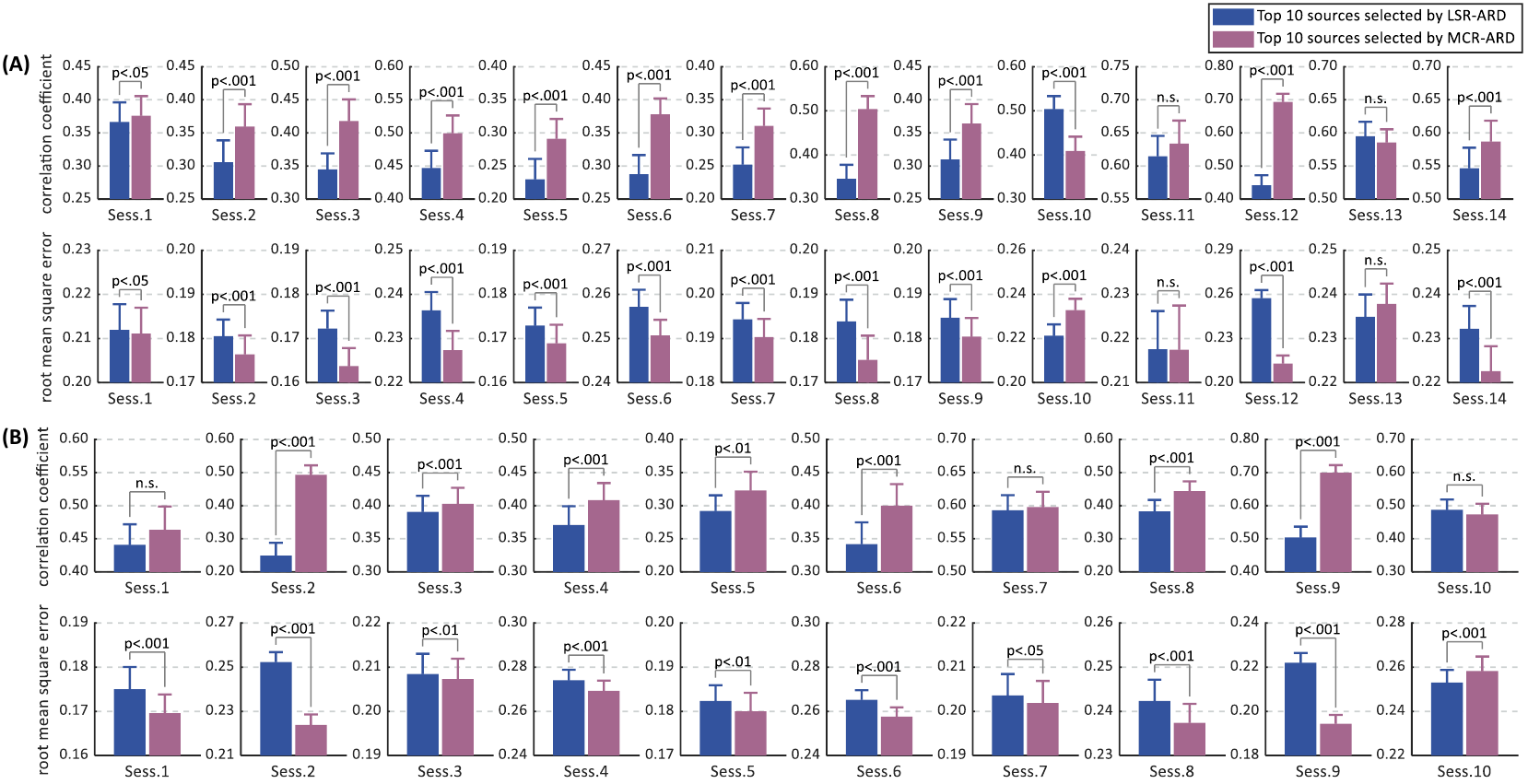}
\caption{Correlation coefficient and root mean squared error for the real-world muscle activity reconstruction dataset employing the top 10 current sources with the largest contributions as the covariates and using the naive LSR algorithm for reconstruction: (A) flexion sessions (B) extension sessions. The results are averaged by repeating the five-fold cross validation for ten repetitions. The error bars denote the 95$\%$ confidence intervals for each result. Paired \textit{t}-test was used to examine the statistically significant difference. ``n.s.'' indicates no significant difference (\textit{p}$>$.05). The top 10 current sources were selected for each session separately.}
\label{figs3}
\end{figure*}

On the other hand, the spatio-temporal feature selection was studied for intra-session decoding. For each session, 50 models were trained (5-fold cross validation $\times$ 10 repetitions). In Fig. \ref{figs1}, as one would perceive, we illustrated the number of selected features averaged across 50 regression models for each session. LSR-ARD and MCR-ARD utilized a stable number of features within each session, while the numbers are obviously different between two algorithms. Concerning the temporal pattern, Fig. \ref{figs2} shows the percentage that each temporal lag was leveraged in 50 models, averaged by all sessions of flexion and extension, respectively. For example, $50\%$ means a temporal lag was used in 25 models. One can observe that, both LSR-ARD and MCR-ARD leveraged the whole 160 ms interval, whereas LSR-ARD had large fluctuations between neighboring lags. On the other hand, regarding spatial feature, the contribution of each source for predicting muscle activity was computed by 
\begin{equation}
\label{equ:spatial}
Contr(s)=\frac{\sum_{t}\left\|w_{s,t}\right\|}{\sum_{s}\sum_{t}\left\|w_{s,t}\right\|}
\end{equation}
where $Contr(s)$ indicates the contribution of $s$-th source. $w_{s,t}$ corresponds to $s$-th source and $t$-th temporal lag. To test which algorithm realized superior source selection, the top 10 sources of the largest contributions for each algorithm were utilized for muscle activity reconstruction using the naive LSR (non-robust and non-sparse). Fig. \ref{figs3} illustrates the result for each session. One can perceive that, even using the same LSR for prediction, the top 10 sources for MCR-ARD achieved higher correlation coefficient and lower root mean squared error with significant difference in most sessions for both flexion and extension. This suggests that, MCR-ARD can select those sources that contain more relevant information for brain decoding than LSR-ARD.

\begin{table*}[t]
	\centering
	\caption{Average decoding performance regrading each muscle task of each subject. The superior results are marked in bold.}
	\label{table_intra}
	\resizebox{0.86\textwidth}{!}{
		\renewcommand{\arraystretch}{1.2}
		\setlength{\tabcolsep}{3.2pt}
		\begin{tabular}{@{}cccccccccccccccc@{}} 
			\toprule
			\hline
			\multicolumn{1}{c|}{\multirow{2}{*}{Performance indicator}} & \multicolumn{1}{c|}{\multirow{2}{*}{Algorithm}} & \multicolumn{3}{c|}{\textit{Subj.1}}& \multicolumn{4}{c|}{\textit{Subj.2}}& \multicolumn{3}{c|}{\textit{Subj.3}}& \multicolumn{4}{c}{\textit{Subj.4}}\\
			\cline{3-16}
			\multicolumn{1}{c|}{\multirow{2}{*}{}} & \multicolumn{1}{c|}{\multirow{2}{*}{}} &
			\textit{FH} & \textit{FL} & \multicolumn{1}{c|}{\textit{EL}} & \textit{FH} & \textit{FL} & \textit{EH} & \multicolumn{1}{c|}{\textit{EL}} & \textit{FH} & \textit{FL} & \multicolumn{1}{c|}{\textit{EL}} & \textit{FH} & \textit{FL} & \textit{EH} & \textit{EL}\\
			\hline
			\multicolumn{1}{c|}{\multirow{2}{*}{Correlation coefficient $\uparrow$}} & \multicolumn{1}{c|}{\multirow{1}{*}{LSR-ARD}} &
			0.6297 & 0.6098 & \multicolumn{1}{c|}{0.5917} & 0.6400 & 0.5182 & 0.7045 & \multicolumn{1}{c|}{0.6559} & 0.5567 & 0.5475 & \multicolumn{1}{c|}{0.6182} & 0.8006 & 0.7332 & 0.7765 & 0.7302\\
			\multicolumn{1}{c|}{\multirow{2}{*}{}} & \multicolumn{1}{c|}{\multirow{1}{*}{MCR-ARD}} &
			\textbf{0.6359} & \textbf{0.6271} & \multicolumn{1}{c|}{\textbf{0.5951}} & \textbf{0.6502} & \textbf{0.5310} & \textbf{0.7149} & \multicolumn{1}{c|}{\textbf{0.6654}} & \textbf{0.5715} & \textbf{0.5560} & \multicolumn{1}{c|}{\textbf{0.6220}} & \textbf{0.8188} & \textbf{0.7536} & \textbf{0.7839} & \textbf{0.7324}\\
			\hline
			\multicolumn{1}{c|}{\multirow{2}{*}{Root mean squared error $\downarrow$}} & \multicolumn{1}{c|}{\multirow{1}{*}{LSR-ARD}} &
			0.1856 & 0.1538 & \multicolumn{1}{c|}{0.1677} & 0.2086 & 0.1657 & 0.2033 & \multicolumn{1}{c|}{0.1605} & 0.1714 & 0.1963 & \multicolumn{1}{c|}{0.2348} & 0.1834 & 0.2068 & 0.1786 & 0.2092\\
			\multicolumn{1}{c|}{\multirow{2}{*}{}} & \multicolumn{1}{c|}{\multirow{1}{*}{MCR-ARD}} &
			\textbf{0.1833} & \textbf{0.1524} & \multicolumn{1}{c|}{\textbf{0.1655}} & \textbf{0.2063} & \textbf{0.1654} & \textbf{0.2014} & \multicolumn{1}{c|}{\textbf{0.1596}} & \textbf{0.1681} & \textbf{0.1911} & \multicolumn{1}{c|}{\textbf{0.2315}} & \textbf{0.1756} & \textbf{0.1923} & \textbf{0.1730} & \textbf{0.2047}\\
			\hline
			\bottomrule
			\multicolumn{16}{r}{\multirow{1}{*}{*\textit{FH}=\textit{FlexHigh}, \textit{FL}=\textit{FlexLow}, \textit{EH}=\textit{ExtHigh}, \textit{EL}=\textit{ExtLow}}}\\
	\end{tabular}}
\end{table*}

\begin{figure*}[t!]
\centering
\includegraphics[width=0.68\textwidth]{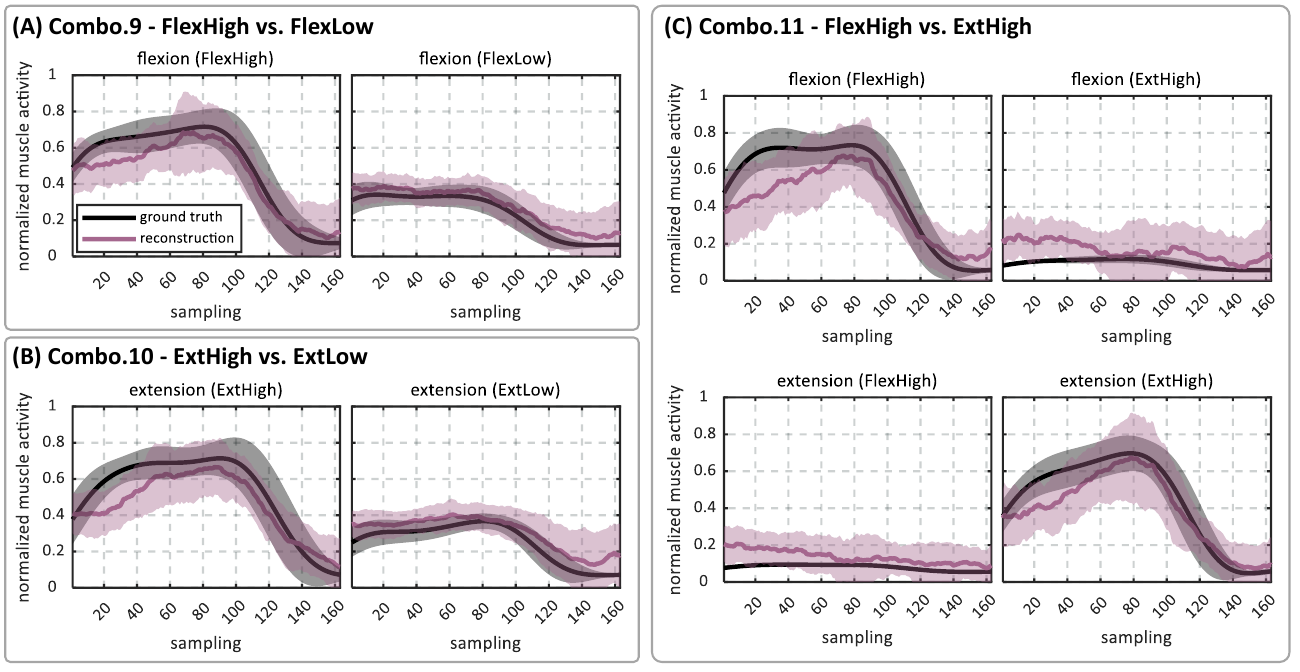}
\caption{Three examples of inter-session decoding by MCR-ARD: (A) Combo.9 (B) Combo.10 (C) Combo.11. The ground truth illustrates the average muscle activity of 50 trials for the corresponding session, while the reconstructions are averaged across 500 testing results of each session (totally 1000 testing results in each combo). The shade area denotes the standard deviation.}
\label{fig_muscle2}
\end{figure*}

\begin{figure*}[t!]
\centering
\includegraphics[width=0.96\textwidth]{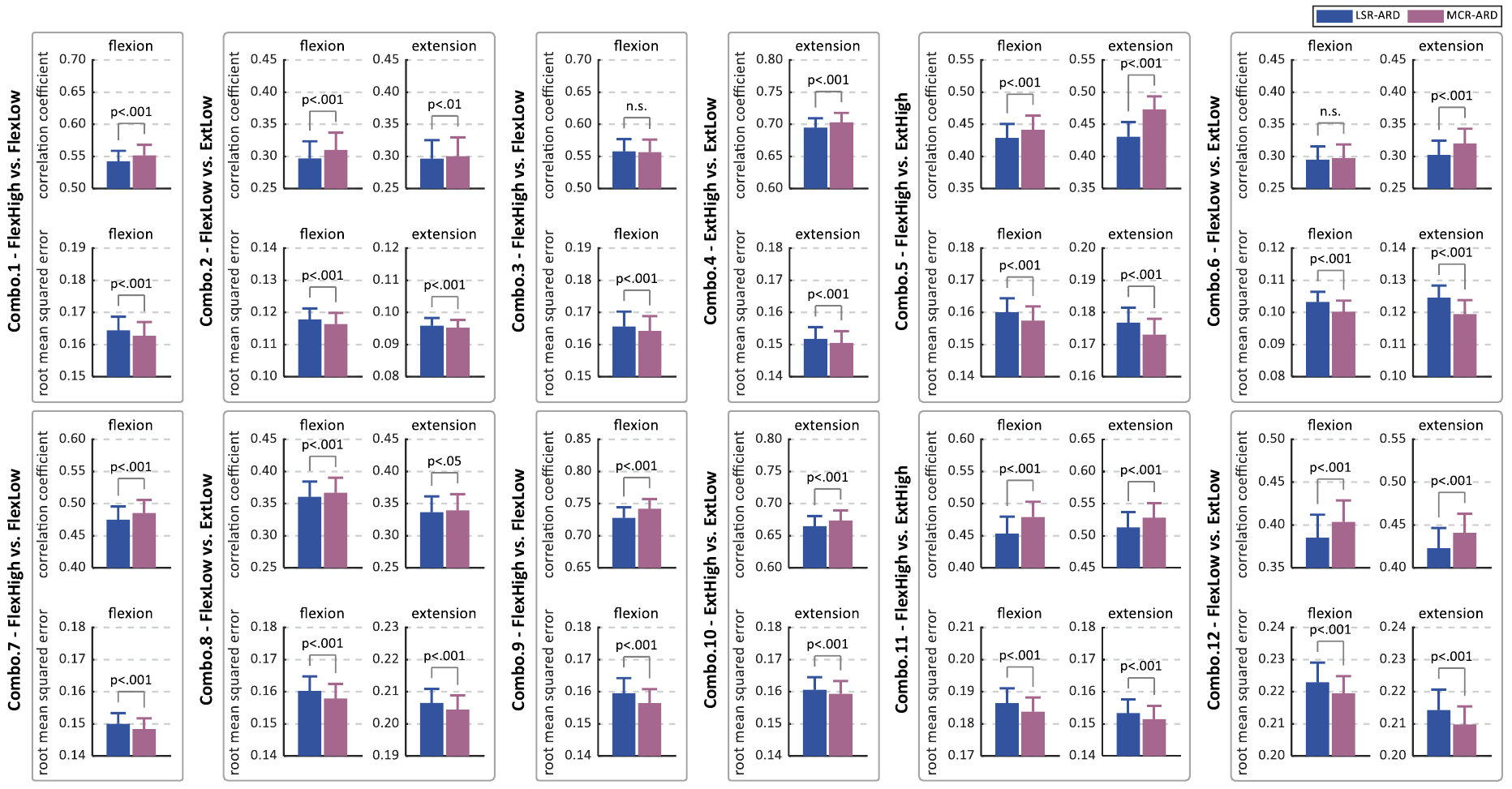}
\caption{Correlation coefficient and root mean squared error for the real-world muscle activity reconstruction dataset by employing the estimated cortical current source signals as the covariates under the inter-session decoding framework. The error bars denote the $95\%$ confidence intervals. Paired $t$-test was used to examine the statistical difference. ``n.s.'' indicates no significant difference (\textit{p}$>$.05).}
\label{fig_muscle_inter_performance}
\end{figure*}

For the inter-session decoding, three examples of identifying different activities (flexion / extension) and forces (high / low) are illustrated in Fig. \ref{fig_muscle2} for the proposed MCR-ARD algorithm. From Fig. \ref{fig_muscle2}(A)(B), one could perceive the obvious compliance of the reconstructed muscle activity for different force for both flexion and extension. Therefore, MCR-ARD could reconstruct muscle activity with a desired force in real-world applications. Moreover, Fig. \ref{fig_muscle2}(C) shows that MCR-ARD can reconstruct the activated and non-activated muscle activity for both flexion and extension simultaneously. This suggests that MCR-ARD could reconstruct the muscle activity for a more comprehensive task. Regarding the quantitative comparison between LSR-ARD and MCR-ARD, Fig.\ref{fig_muscle_inter_performance} illustrates the inter-session decoding result for each combination. One observes that, except the correlation coefficients of flexion for Combo.3 and Combo.6, the proposed MCR-ARD outperformed the conventional LSR-ARD in every condition with statistically significant difference.

\subsection{Muscle Activity Reconstruction by EEG}
\label{sec:eeg}

\begin{figure*}[t!]
	\centering
	\includegraphics[width=0.92\textwidth]{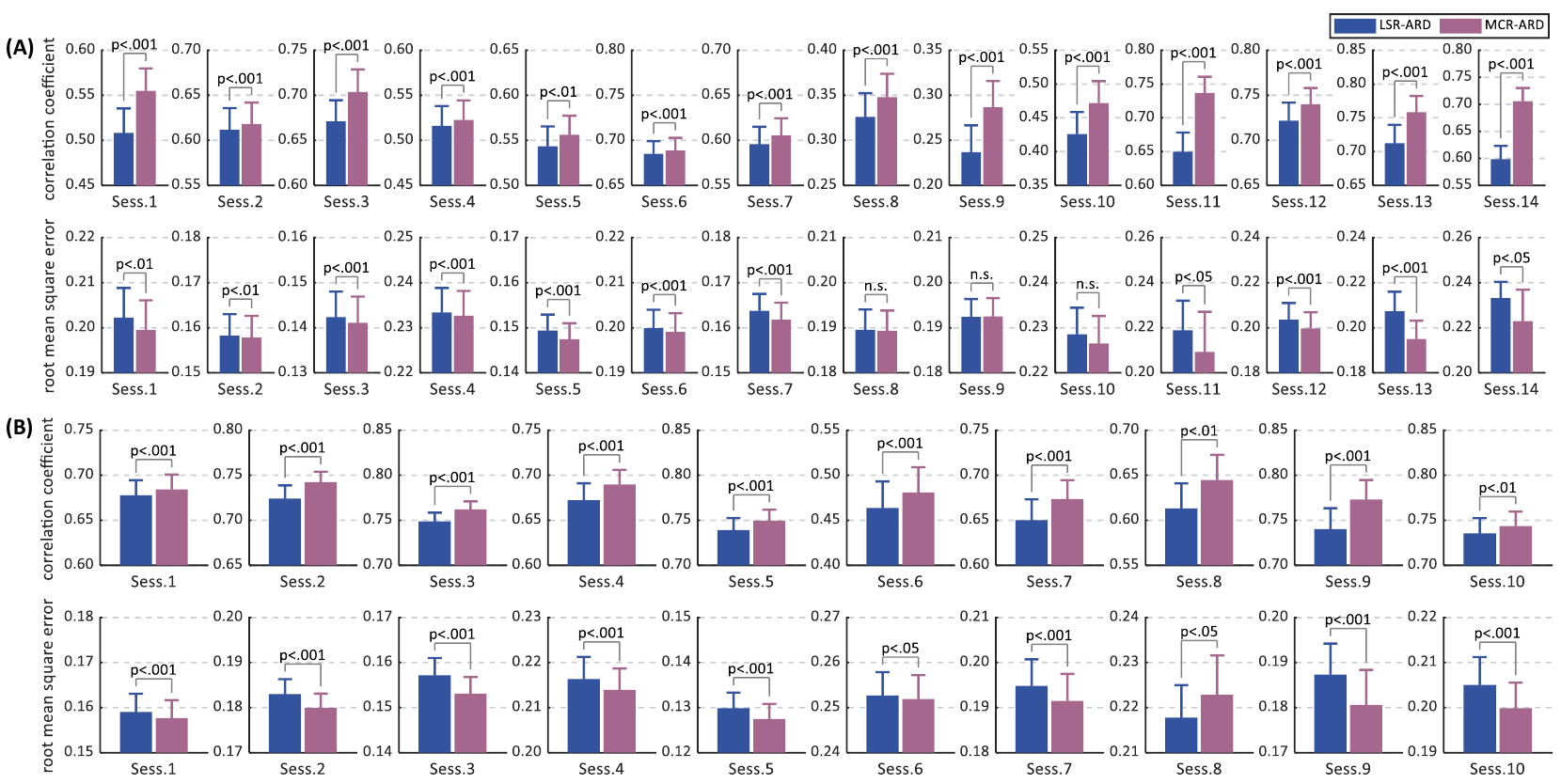}
	\caption{Correlation coefficient and root mean squared error for the real-world muscle activity reconstruction dataset by employing the preprocessed 30-channel EEG signals as the covariates:(A) flexion sessions (B) extension sessions. The results are averaged by repeating the five-fold cross validation for ten repetitions. The error bars denote the $95\%$ confidence intervals for each result. Paired $t$-test was conducted to examine the statistically significant difference. ``n.s.'' indicates no significant difference (\textit{p}$>$.05).}
	\label{fig_eeg}
\end{figure*}

In EEG-based intra-session decoding, Fig. \ref{fig_eeg} illustrates the reconstruction result for each session. One can observe, similar to before, that the proposed MCR-ARD achieved significantly superior prediction than LSR-ARD on the majority of sessions. Specifically, for 14 flexion sessions, MCR-ARD outperformed LSR-ARD with significant difference for all sessions regarding correlation coefficient, and for 11 sessions regarding root mean squared error. For 10 extension sessions, MCR-ARD achieved significantly higher correlation coefficient and lower root mean squared error for all sessions and 9 sessions, respectively. This further authenticates the admirable competence of MCR-ARD for decoding sophisticated activity using noisy brain recording.

\section{Discussion}
\label{sec:dis}

This study proposed a novel robust sparse regression method named as MCR-ARD, which provides a powerful tool for brain activity decoding. Although the conference paper for this study \cite{li2023adaptive} has introduced the theoretical derivation for the algorithm and used a synthetic example with noisy output for evaluation, it remained unclear whether MCR-ARD could realize superior neural decoding performance in a real-world application. Since the noisy brain recording is used as the covariate for decoding, the present paper first employed a synthetic dataset with noisy covariate, which is a more realistic problem setting for the real- world brain activity decoding. To evaluate the proposed MCR- ARD algorithm with real brain datasets, this paper further used the muscle activity reconstruction scenario with current source and EEG signal. The experimental results in this paper suggest that, our proposed MCR-ARD algorithm could provide a better substitute for brain signal decoding instead of the conventional sparse Bayesian learning method. Furthermore, because MCR-ARD was proposed in a primary perspective, i.e. the likelihood function, it can be conveniently employed with other Bayesian techniques to realize further improvement. For example, block-sparse Bayesian learning (BSBL) has been employed for EEG source estimation \cite{qu2022nonnegative}, since block-sparsity could better model the EEG signals which are more likely to generate from locally synchronized neural issues. Integrating the proposed likelihood function (\ref{equ:mcc_likelihood_function}) with BSBL is expected to be more adequate since the proposed likelihood function will help solve the intractable problem of EEG recording noises. The proposed robust sparse regression algorithm is in particular applicable for the scenario where it is difficult to collect a large amount of brain data, e.g., collecting both brain and muscle data meanwhile in the muscle activity reconstruction task. Complicated neural network, such as the convolutional neural network (CNN) and the long short-term memory (LSTM), will be difficult to train for such a case, while our proposed method could provide a practical approach. MCR-ARD could be utilized to structure a more advanced BCI system for continuous motion prediction. For example, muscle activity can be predicted from brain recordings by MCR-ARD, which could be further leveraged to control the prosthetic hand in real time for motion assistance \cite{qin2021multi,qin2022cw}. MCR-ARD will help improve the BCI systems to realize better motion reproduction and rehabilitation.

Robustness and sparseness are two crucial natures for many machine learning tasks. Existing robust sparse methods mainly focus on regularizing a robust objective function by a sparsity-promoting regularization term, e.g., $L_1$-regularized MCC \cite{he2010maximum,ma2015maximum}. However, these methods have two hyper-parameters, one controlling robustness and another one controlling sparseness. For example, $L_1$-regularized MCC tunes kernel bandwidth for MCC and regularization weight for $L_1$-norm. This usually asks for a grid search to find the optimal configurations for the two hyper-parameters, which can be considerably time-consuming in practice. To reduce hyper-parameter, we leveraged the ARD method which can realize adaptive sparseness. However, ARD does not have a factorable regularization form \cite{wipf2007new}. Hence, we cannot combine MCC and ARD directly like existing methods. To address this obstacle, we innovatively exposed the inherent error assumption in MCC with an explicit form $\mathcal{C}(\epsilon|0,h)$. Thus we can integrate robust MCC with ARD in a Bayesian regime. To the best of our knowledge, this was the first time to explain MCC from a Bayesian viewpoint. Compared to $L_1$-regularized MCR, our proposed MCR-ARD can realize better performance for the noisy and high-dimensional regression, as shown in the preliminary conference paper \cite{li2023adaptive}. More crucially, MCR-ARD only has one hyperparameter to tune, i.e. the kernel bandwidth, which will greatly facilitate its applications to real-world tasks. The reason why we did not assess $L_1$-regularized MCR in this paper was it needed unacceptable time to finish the grid search to select two hyper-parameters. For example, if we select both kernel bandwidth and regularization weight from 30 candidate values, we need to train the model with 900 combinations. By comparison, MCR-ARD only needs a line search to determine $h$ from 30 candidates. As a result, MCR-ARD greatly reduces the time to select the hyperparameter, which makes it evidently easier to use in the real-world brain activity decoding scenario.

The core insight of this research is that our proposed deviant error assumption $\mathcal{C}(\epsilon|0,h)$ realizes much better robustness than the conventional Gaussian distribution regarding the utilization for the likelihood function. The decisive property of $\mathcal{C}(\epsilon|0,h)$, as mentioned in Section \ref{sec:mcc_noise}, is $\lim_{\epsilon\rightarrow\infty}\mathcal{C}(\epsilon|0,h)=1$, while the generic distributions will converge to $0$. We desire to argue that, it is this difference that greatly ameliorates the robustness. To be specific, if a dataset contains adverse noises, they would result in training errors with arbitrarily large amplitudes. Thus, the probability of a large error occurring is not zero, conflicting with the traditional error assumptions. In contrast, our deviant $\mathcal{C}(\epsilon|0,h)$ assumes that any arbitrarily large error might happen with a non-zero probability, which is more sensible for a noisy scenario. On the other hand, by tradition, improper distribution is only allowed for the prior distribution for Bayesian learning \cite{gelman1995bayesian}. This study employed the deviant assumption $\mathcal{C}(\epsilon|0,h)$ for the likelihood, and thus realized significant improvements. The integration of MCC into the Bayesian learning framework will be further scrutinized in our future works, with the motivations of recent studies that have made effort to establish a theoretical guarantee for the non-normalized likelihood models \cite{giummole2019objective,matsuda2021information}.

We would also like to discuss about the only hyperparameter in our MCR-ARD, the kernel bandwidth $h$. It was determined by utilizing cross validation (Section \ref{sec:toy_des}) or an independent validation set (Section \ref{sec:eval2} and \ref{sec:eval3}). As mentioned before, $h$ was regarded as a fixed parameter in our algorithm. It would be an interesting question whether we can make $h$ as a random variable and incorporate it in the Bayesian inference processes, so that $h$ could be determined automatically to realize adaptive robustness. This topic will be investigated in our future studies.

\begin{figure*}[t]
\centering
\includegraphics[width=0.76\textwidth]{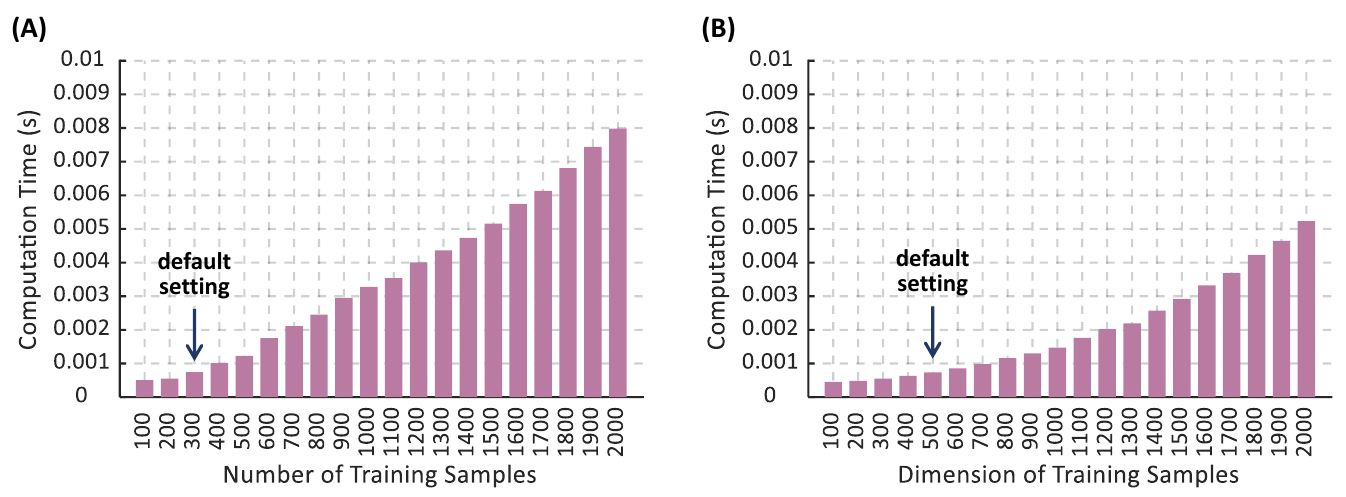}
\caption{Average computation time for one free energy minimization iteration under different dataset scales: (A) under different numbers of training samples (the dimension is fixed to be 500). (B) under different dimensions of training samples (the number of training samples is fixed to be 300). The number of the relevant features is fixed to be 30 as in the original synthetic example.}
\label{figs4}
\end{figure*}

In addition, we also investigated how well MCR-ARD could be generalized for datasets with different scales. We leveraged the synthetic dataset, as mentioned in Section \ref{sec:toy_des}, and used different numbers of samples and different dimensions to study how the training time was affected. Fig. \ref{figs4} shows the average computation time for one iteration of free energy minimization (step 4-step 10 in Algorithm \ref{cslr}) obtained by MATLAB R2023b (MathWorks Inc.) on Intel Core i9-13900H CPU, 32GB RAM, and Windows 11 OS. One could find that the computation time increased linearly with respect to the number of samples, since both \textit{w-step} (\ref{equ:w_fp}) and the calculation of negative Hessian matrix (\ref{equ:negative_hessian}) have linear complexity in regard to $N$. On the other hand, MCR-ARD has a polynomial complexity regrading dimension $D$, resulting from \textit{w-step} (\ref{equ:w_fp}) of $\mathcal{O}(ND^2+D^3)$, and Hessian matrix inverse of $\mathcal{O}(D^3)$. This indicates that MCR-ARD could be well applied for various scenarios with different data scales.

Finally, for the proposed MCR-ARD algorithm in this study, the limitations are discussed as follows. The critical difference between MCR-ARD and conventional LSR-ARD is that, LSR-ARD utilizes the Gaussian error assumption that may be overly idealistic whereas MCR-ARD employs the deviant assumption $\mathcal{C}(\epsilon|0,h)$ in the likelihood function with heavier tail. However, it will be hard to guarantee that $\mathcal{C}(\epsilon|0,h)$ is always the optimal choice for the error assumption in every context. For example, from Fig. \ref{fig_mcc_dist}, one can observe that $\mathcal{C}(\epsilon|0,h)$ cannot approximate the Gaussian distribution naturally. As a result, $\mathcal{C}(\epsilon|0,h)$ would be less adequate and lead to a biased solution for the problems with a Gaussian error distribution. On the other hand, $\mathcal{C}(\epsilon|0,h)$ with one single peak is not appropriate for a more complicated error distribution, e.g., the multi-peak error distribution. Using a multi-peak model may further help improve the brain activity decoding accuracy \cite{chen2018common,li2021restricted}. Hence, it would be an interesting future work to investigate an improved error assumption which has a better flexibility while maintains the desirable robustness.

\section{Conclusion}
\label{sec:con}
To realize robust sparse regression, this research first derived an explicit error assumption $\mathcal{C}(\epsilon|0,h)$ inherent in robust MCR, and utilized it to construct a robust likelihood function. Using ARD technique, MCR was integrated with the sparse Bayesian learning framework. MCR-ARD was systematically evaluated by a synthetic dataset with noisy input and a real-world muscle activity reconstruction task with two different brain modalities. The experimental results show that our proposed algorithm can largely ameliorate the robustness for sparse Bayesian learning. The proposed \textit{\textbf{sparse Bayesian correntropy learning}} method, including our previous work \cite{li2023correntropy}, provides a powerful tool for the real-world brain activity decoding tasks to realize advanced brain-computer interface technology.

\bibliographystyle{IEEEtran}
\bibliography{bibib}
\end{document}